
\input epsf

\def\oldstyle#1{$\mit #1$}
\def\dalemb{
	\hbox{\vrule\vbox to 6pt{\hrule width 6pt
	\vfill\hrule}\vrule}\,}%

\font\cmbxXIVf=cmbx10 at 14pt
\font\cmbxXIIIf=cmbx10 at 13pt

\font\cmbxXII=cmbx10 at 12pt

\font\tenitbf=cmbxti10 at 10pt
\font\sevenitbf=cmbxti10 at 7pt
\font\fiveitbf=cmbxti10 at 5pt
\font\itbfXIII=cmbxti10 at 13pt
\font\itbfXIV=cmbxti10 at 14pt
\font\itbfX=cmbxti10 at 10pt
\font\itbfXII=cmbxti10 at 12pt

\newfam\itbf
\textfont\itbf=\tenitbf
\scriptfont\itbf=\sevenitbf
\scriptscriptfont\itbf=\fiveitbf
\def\ibf{\fam\itbf\tenitbf}

\newfam\itbfXIIIfont
\textfont\itbfXIIIfont=\itbfXIII
\scriptfont\itbfXIIIfont=\itbfXII
\scriptscriptfont\itbfXIIIfont=\itbfX
\def\cmbxXIII{\fam\itbfXIIIfont\cmbxXIIIf}

\newfam\itbfXIVfont
\textfont\itbfXIVfont=\itbfXIV
\scriptfont\itbfXIVfont=\itbfXII
\scriptscriptfont\itbfXIVfont=\itbfX
\def\cmbxXIV{\fam\itbfXIVfont\cmbxXIVf}

\expandafter\ifx\csname amssym.def\endcsname\relax \else\endinput\fi
\expandafter\edef\csname amssym.def\endcsname{%
       \catcode`\noexpand\@=\the\catcode`\@\space}
\catcode`\@=11

\def\undefine#1{\let#1\undefined}
\def\newsymbol#1#2#3#4#5{\let\next@\relax
 \ifnum#2=\@ne\let\next@\msafam@\else
 \ifnum#2=\tw@\let\next@\msbfam@\fi\fi
 \mathchardef#1="#3\next@#4#5}
\def\mathhexbox@#1#2#3{\relax
 \ifmmode\mathpalette{}{\m@th\mathchar"#1#2#3}%
 \else\leavevmode\hbox{$\m@th\mathchar"#1#2#3$}\fi}
\def\hexnumber@#1{\ifcase#1 0\or 1\or 2\or 3\or 4\or 5\or 6\or 7\or 8\or
 9\or A\or B\or C\or D\or E\or F\fi}

\font\tenmsa=msam10
\font\sevenmsa=msam7
\font\fivemsa=msam5
\newfam\msafam
\textfont\msafam=\tenmsa
\scriptfont\msafam=\sevenmsa
\scriptscriptfont\msafam=\fivemsa
\edef\msafam@{\hexnumber@\msafam}
\mathchardef\dabar@"0\msafam@39
\def\dashrightarrow{\mathrel{\dabar@\dabar@\mathchar"0\msafam@4B}}
\def\dashleftarrow{\mathrel{\mathchar"0\msafam@4C\dabar@\dabar@}}

\def\ulcorner{\delimiter"4\msafam@70\msafam@70 }
\def\urcorner{\delimiter"5\msafam@71\msafam@71 }
\def\llcorner{\delimiter"4\msafam@78\msafam@78 }
\def\lrcorner{\delimiter"5\msafam@79\msafam@79 }
\def\yen{{\mathhexbox@\msafam@55 }}
\def\checkmark{{\mathhexbox@\msafam@58 }}
\def\circledR{{\mathhexbox@\msafam@72 }}
\def\maltese{{\mathhexbox@\msafam@7A }}

\font\tenmsb=msbm10
\font\sevenmsb=msbm7
\font\fivemsb=msbm5
\newfam\msbfam
\textfont\msbfam=\tenmsb
\scriptfont\msbfam=\sevenmsb
\scriptscriptfont\msbfam=\fivemsb
\edef\msbfam@{\hexnumber@\msbfam}

\def\widehat#1{\setbox\z@\hbox{$\m@th#1$}%
 \ifdim\wd\z@>\tw@ em\mathaccent"0\msbfam@5B{#1}%
 \else\mathaccent"0362{#1}\fi}
\def\widetilde#1{\setbox\z@\hbox{$\m@th#1$}%
 \ifdim\wd\z@>\tw@ em\mathaccent"0\msbfam@5D{#1}%
 \else\mathaccent"0365{#1}\fi}
\font\teneufm=eufm10
\font\seveneufm=eufm7
\font\fiveeufm=eufm5
\newfam\eufmfam
\textfont\eufmfam=\teneufm
\scriptfont\eufmfam=\seveneufm
\scriptscriptfont\eufmfam=\fiveeufm

\csname amssym.def\endcsname

\expandafter\ifx\csname pre amssym.tex at\endcsname\relax \else \endinput\fi
\expandafter\chardef\csname pre amssym.tex at\endcsname=\the\catcode`\@
\catcode`\@=11
\newsymbol\boxdot 1200
\newsymbol\boxplus 1201
\newsymbol\boxtimes 1202
\newsymbol\square 1003
\newsymbol\blacksquare 1004
\newsymbol\centerdot 1205
\newsymbol\lozenge 1006
\newsymbol\blacklozenge 1007
\newsymbol\circlearrowright 1308
\newsymbol\circlearrowleft 1309
\undefine\rightleftharpoons
\newsymbol\rightleftharpoons 130A
\newsymbol\leftrightharpoons 130B
\newsymbol\boxminus 120C
\newsymbol\Vdash 130D
\newsymbol\Vvdash 130E
\newsymbol\vDash 130F
\newsymbol\twoheadrightarrow 1310
\newsymbol\twoheadleftarrow 1311
\newsymbol\leftleftarrows 1312
\newsymbol\rightrightarrows 1313
\newsymbol\upuparrows 1314
\newsymbol\downdownarrows 1315
\newsymbol\upharpoonright 1316
 
\newsymbol\downharpoonright 1317
\newsymbol\upharpoonleft 1318
\newsymbol\downharpoonleft 1319
\newsymbol\rightarrowtail 131A
\newsymbol\leftarrowtail 131B
\newsymbol\leftrightarrows 131C
\newsymbol\rightleftarrows 131D
\newsymbol\Lsh 131E
\newsymbol\Rsh 131F
\newsymbol\rightsquigarrow 1320
\newsymbol\leftrightsquigarrow 1321
\newsymbol\looparrowleft 1322
\newsymbol\looparrowright 1323
\newsymbol\circeq 1324
\newsymbol\succsim 1325
\newsymbol\gtrsim 1326
\newsymbol\gtrapprox 1327
\newsymbol\multimap 1328
\newsymbol\therefore 1329
\newsymbol\because 132A
\newsymbol\doteqdot 132B
 
\newsymbol\triangleq 132C
\newsymbol\precsim 132D
\newsymbol\lesssim 132E
\newsymbol\lessapprox 132F
\newsymbol\eqslantless 1330
\newsymbol\eqslantgtr 1331
\newsymbol\curlyeqprec 1332
\newsymbol\curlyeqsucc 1333
\newsymbol\preccurlyeq 1334
\newsymbol\leqq 1335
\newsymbol\leqslant 1336
\newsymbol\lessgtr 1337
\newsymbol\backprime 1038
\newsymbol\risingdotseq 133A
\newsymbol\fallingdotseq 133B
\newsymbol\succcurlyeq 133C
\newsymbol\geqq 133D
\newsymbol\geqslant 133E
\newsymbol\gtrless 133F
\newsymbol\sqsubset 1340
\newsymbol\sqsupset 1341
\newsymbol\vartriangleright 1342
\newsymbol\vartriangleleft 1343
\newsymbol\trianglerighteq 1344
\newsymbol\trianglelefteq 1345
\newsymbol\bigstar 1046
\newsymbol\between 1347
\newsymbol\blacktriangledown 1048
\newsymbol\blacktriangleright 1349
\newsymbol\blacktriangleleft 134A
\newsymbol\vartriangle 134D
\newsymbol\blacktriangle 104E
\newsymbol\triangledown 104F
\newsymbol\eqcirc 1350
\newsymbol\lesseqgtr 1351
\newsymbol\gtreqless 1352
\newsymbol\lesseqqgtr 1353
\newsymbol\gtreqqless 1354
\newsymbol\Rrightarrow 1356
\newsymbol\Lleftarrow 1357
\newsymbol\veebar 1259
\newsymbol\barwedge 125A
\newsymbol\doublebarwedge 125B
\undefine\angle
\newsymbol\angle 105C
\newsymbol\measuredangle 105D
\newsymbol\sphericalangle 105E
\newsymbol\varpropto 135F
\newsymbol\smallsmile 1360
\newsymbol\smallfrown 1361
\newsymbol\Subset 1362
\newsymbol\Supset 1363
\newsymbol\Cup 1264
 
\newsymbol\Cap 1265
 
\newsymbol\curlywedge 1266
\newsymbol\curlyvee 1267
\newsymbol\leftthreetimes 1268
\newsymbol\rightthreetimes 1269
\newsymbol\subseteqq 136A
\newsymbol\supseteqq 136B
\newsymbol\bumpeq 136C
\newsymbol\Bumpeq 136D
\newsymbol\lll 136E
 
\newsymbol\ggg 136F
 
\newsymbol\circledS 1073
\newsymbol\pitchfork 1374
\newsymbol\dotplus 1275
\newsymbol\backsim 1376
\newsymbol\backsimeq 1377
\newsymbol\complement 107B
\newsymbol\intercal 127C
\newsymbol\circledcirc 127D
\newsymbol\circledast 127E
\newsymbol\circleddash 127F
\newsymbol\lvertneqq 2300
\newsymbol\gvertneqq 2301
\newsymbol\nleq 2302
\newsymbol\ngeq 2303
\newsymbol\nless 2304
\newsymbol\ngtr 2305
\newsymbol\nprec 2306
\newsymbol\nsucc 2307
\newsymbol\lneqq 2308
\newsymbol\gneqq 2309
\newsymbol\nleqslant 230A
\newsymbol\ngeqslant 230B
\newsymbol\lneq 230C
\newsymbol\gneq 230D
\newsymbol\npreceq 230E
\newsymbol\nsucceq 230F
\newsymbol\precnsim 2310
\newsymbol\succnsim 2311
\newsymbol\lnsim 2312
\newsymbol\gnsim 2313
\newsymbol\nleqq 2314
\newsymbol\ngeqq 2315
\newsymbol\precneqq 2316
\newsymbol\succneqq 2317
\newsymbol\precnapprox 2318
\newsymbol\succnapprox 2319
\newsymbol\lnapprox 231A
\newsymbol\gnapprox 231B
\newsymbol\nsim 231C
\newsymbol\ncong 231D
\newsymbol\diagup 231E
\newsymbol\diagdown 231F
\newsymbol\varsubsetneq 2320
\newsymbol\varsupsetneq 2321
\newsymbol\nsubseteqq 2322
\newsymbol\nsupseteqq 2323
\newsymbol\subsetneqq 2324
\newsymbol\supsetneqq 2325
\newsymbol\varsubsetneqq 2326
\newsymbol\varsupsetneqq 2327
\newsymbol\subsetneq 2328
\newsymbol\supsetneq 2329
\newsymbol\nsubseteq 232A
\newsymbol\nsupseteq 232B
\newsymbol\nparallel 232C
\newsymbol\nmid 232D
\newsymbol\nshortmid 232E
\newsymbol\nshortparallel 232F
\newsymbol\nvdash 2330
\newsymbol\nVdash 2331
\newsymbol\nvDash 2332
\newsymbol\nVDash 2333
\newsymbol\ntrianglerighteq 2334
\newsymbol\ntrianglelefteq 2335
\newsymbol\ntriangleleft 2336
\newsymbol\ntriangleright 2337
\newsymbol\nleftarrow 2338
\newsymbol\nrightarrow 2339
\newsymbol\nLeftarrow 233A
\newsymbol\nRightarrow 233B
\newsymbol\nLeftrightarrow 233C
\newsymbol\nleftrightarrow 233D
\newsymbol\divideontimes 223E
\newsymbol\varnothing 203F
\newsymbol\nexists 2040
\newsymbol\Finv 2060
\newsymbol\Game 2061
\newsymbol\mho 2066
\newsymbol\eth 2067
\newsymbol\eqsim 2368
\newsymbol\beth 2069
\newsymbol\gimel 206A
\newsymbol\daleth 206B
\newsymbol\lessdot 236C
\newsymbol\gtrdot 236D
\newsymbol\ltimes 226E
\newsymbol\rtimes 226F
\newsymbol\shortmid 2370
\newsymbol\shortparallel 2371
\newsymbol\smallsetminus 2272
\newsymbol\thicksim 2373
\newsymbol\thickapprox 2374
\newsymbol\approxeq 2375
\newsymbol\succapprox 2376
\newsymbol\precapprox 2377
\newsymbol\curvearrowleft 2378
\newsymbol\curvearrowright 2379
\newsymbol\digamma 207A
\newsymbol\varkappa 207B
\newsymbol\Bbbk 207C
\newsymbol\hslash 207D
\undefine\hbar
\newsymbol\hbar 207E
\newsymbol\backepsilon 237F
\catcode`\@=\csname pre amssym.tex at\endcsname

\font\teneurm=eurm10
\font\seveneurm=eurm7
\font\fiveeurm=eurm5
\newfam\eurmfam
\textfont\eurmfam=\teneurm
\scriptfont\eurmfam=\seveneurm
\scriptscriptfont\eurmfam=\fiveeurm
\def\eulr{\fam\eurmfam\teneurm}

\font\teneusm=eusm10
\font\seveneusm=eusm7
\font\fiveeusm=eusm5
\newfam\eusmfam
\textfont\eusmfam=\teneusm
\scriptfont\eusmfam=\seveneusm
\scriptscriptfont\eusmfam=\fiveeusm

\font\tenmsb=msbm10
\textfont\msbfam=\tenmsb

\font\teneufm=eufm10
\textfont\eufmfam=\teneufm




\def\numero{\hbox{n${}^\circ$} }


\catcode`\@=11
\catcode`\,=11
\def\bibdot{\leaders\hbox to 5mm{\hfil.\hfil}\hfill}%

\newcount\c@chapter
\newcount\c@section
\newcount\c@subsection
\newcount\c@annexe
\newcount\c@asection
\newcount\c@asubsection
\newcount\cfootnot
\newcount\cfigure
\newcount\ctableau
\newcount\cno
\newcount\c@count
\newcount\c@loop
\newcount\c@tot
\newcount\c@min
\newcount\c@Min
\newcount\numvoir
\newcount\@t
\newcount\@f
\newcount\@tt
\newcount\@ff
\newcount\@fff
\newcount\c@biblio

\newif\ifend

\newtoks\repere
\newtoks\author
\newtoks\title
\newtoks\ref


\def\chap{chapitre}
\def\ann{annexe}

\def\chapter#1{\global\advance\c@chapter by 1%
\relax
\relax\c@section=0\relax\c@subsection=0%
\relax
\immediate\write4{part \the\c@chapter}%
\vskip 2cm\goodbreak%
\def\typepartie{chapitre}%
\setbox1=\hbox{{\cmbxXIV \uppercase\expandafter%
{\romannumeral\the\c@chapter}.\hskip 5mm #1}}%
\centerline{\box1}
\nobreak\vskip 0.5cm\nobreak}%

\def\section#1{\global\advance\c@section by 1%
\relax\c@subsection=0%
\vskip 5mm plus 3mm minus 1mm \penalty -500%
\setbox1=\hbox{{\cmbxXIII #1}}
{\cmbxXIII \the\c@chapter.\the\c@section\hskip 4mm \box1}%
\nobreak%
\vskip 2mm plus 1mm minus 0.5mm%
\nobreak}%

\def\asection#1{\global\advance\c@asection by 1%
\relax\c@asubsection=0%
\vskip 5mm plus 3mm minus 1mm\penalty -200%
\setbox1=\hbox{{\cmbxXIII #1}}
{\cmbxXIII \the\c@asection\hskip 4mm \box1}%
\penalty 2000%
\vskip 2mm plus 1mm minus 0.5mm%
\penalty 2000}%

\def\subsection#1{\global\advance\c@subsection by 1%
\vskip 3mm plus 1mm minus 1mm\penalty -200%
\setbox1=\hbox{{\cmbxXII #1}}
{\cmbxXII \the\c@chapter.\the\c@section.\the\c@subsection\hskip 4mm \box1}%
\penalty 2000%
\vskip 1.5mm plus 1mm minus 0.5mm%
\penalty 2000}%

\def\asubsection#1{\global\advance\c@asubsection by 1%
\vskip 3mm plus 1mm minus 1mm\penalty -200%
\setbox1=\hbox{{\cmbxXII #1}}
{\cmbxXII \the\c@asection.\the\c@asubsection\hskip 4mm \box1}%
\penalty 2000%
\vskip 1.5mm plus 1mm minus 0.5mm%
\penalty 2000}%

\def\annexe#1{\global\advance\c@annexe by 1%
\relax
\c@asection=0\relax\c@asubsection=0%
\relax
\immediate\write4{appendix\the\c@annexe}%
\vskip 0.6cm\goodbreak%
\def\typepartie{annexe}%
\setbox1=\hbox{\cmbxXIV Appendix\hskip 4mm\the\c@annexe}%
\setbox3=\hbox{{\cmbxXIV \ :\  #1}}
\centerline{\box1\box3}%
\nobreak\vskip 0.6cm\nobreak}%

\def\footnot#1{\global\advance\cfootnot by 1%
\footnote{$^{\scriptscriptstyle (\the\cfootnot)}$}{#1}}%

\def\footnoterule{\kern -3pt \hrule width 2truein \kern 2.4pt}

\def\ie{\hbox{i.e.}\ }


\def\beginbiblio{\immediate\openout3=tbiblio%
        \immediate\write3{\null\vskip 1cm}%
        \immediate\write3{\noexpand\centerline%
         {\hbox{{\cmbxXIV
	 REFERENCES}}}}%
        \immediate\write3{\vskip 0.8cm}%
        \c@biblio=0\relax%
	\immediate\openout2=rbiblio%
	\immediate\closeout2%
	\immediate\openout7=Biblio.tex
	}%
\def\bibmodif#1#2#3#4{\immediate\write7{\noexpand #1}\relax
		\immediate\write7{\noexpand #2}\relax
		\immediate\write7{\noexpand #3}\relax
		\immediate\write7{\noexpand #4}\relax}
\def\bibitem#1#2#3#4{\global\advance\c@biblio by 1\relax%
        \immediate\write3{\vskip 2mm}%
        \immediate\write3{$[\the\c@biblio]$\hskip 3mm {\sl #2}
                           \hskip 3mm #4}%
        \append{#1}{\the\c@biblio}
        }%
\def\bibtitre#1{\immediate\write3{\vskip 4mm}%
         \immediate\write3{\vtop{%
         \hbox{{\bf #1}} \kern 1pt\hrule\kern 1pt\hrule}}}%
\def\endbiblio{\immediate\closeout7}%

\def\append#1#2{\@fff=0%
		\immediate\openout5=tempo%
		\immediate\openin8=rbiblio%
		\loop\ifeof8{}\else%
		\immediate\read8to\toto%
		\immediate\write5{\toto}\fi%
		\ifeof8\@fff=1\fi%
		\ifnum\@fff=0\relax\repeat\immediate\closein8%
		\@fff=0%
		\immediate\write5{#1}\relax%
		\immediate\write5{#2}\relax%
		\immediate\closeout5%
 		\immediate\openin5=tempo%
		\immediate\openout8=rbiblio%
		\loop\ifeof5{}\else%
		\immediate\read5to\toto%
		\immediate\write8{\toto}\fi%
		\ifeof5\@fff=1\fi%
		\ifnum\@fff=0\repeat\immediate\closein5%
		\immediate\closeout8}%

\def\findsubr#1{\immediate\openout5=tempo\relax%
		\immediate\write5{#1}%
		\immediate\closeout5%
		\immediate\openin5=tempo\relax%
		\immediate\read5to\toto%
		\immediate\closein5%
		\immediate\openin2=rbiblio\relax%
		\global\@t=0\relax%
		\global\@f=0\relax%
		\endtrue%
		\loop\ifeof2{}\else%
		\immediate\read2to\titi\relax%
		\immediate\read2to\temp\relax%
		\immediate\read2to\numero\relax%
		\immediate\read2to\tutu\relax%
		\ifx\toto\temp\relax%
		\immediate\write9{\numero}\global\@t=1\relax%
		\fi%
		\fi%
		\ifeof2\@f=1\relax\fi%
		\ifnum\@t=0\relax\ifnum\numero<\c@biblio\relax%
		\endtrue\else\endfalse\fi\else\endfalse\fi%
		\ifend\repeat%
		\immediate\closein2}%

\def\findsub#1{\immediate\openout5=tempo\relax%
		\immediate\write5{#1}%
		\immediate\closeout5%
		\immediate\openin5=tempo\relax%
		\immediate\read5to\toto%
		\immediate\closein5%
		\immediate\openin2=rbiblio\relax%
		\global\@t=0\relax%
		\global\@f=0\relax%
		\endtrue%
		\loop\ifeof2{}\else%
		\immediate\read2to\titi\relax%
		\immediate\read2to\temp\relax%
		\immediate\read2to\numero\relax%
		\immediate\read2to\tutu\relax%
		\ifx\toto\temp\relax\numero\global\@t=1\relax%
		\fi%
		\fi%
		\ifeof2\@f=1\relax\fi%
		\ifnum\@t=0\relax\ifnum\numero<\c@biblio\relax%
		\endtrue\else\endfalse\fi\else\endfalse\fi%
		\ifend\repeat%
		\immediate\closein2}%

\def\findprinc#1{\immediate\openout5=tempo\relax%
		\immediate\write5{#1}%
		\immediate\closeout5%
		\immediate\openin5=tempo\relax%
		\immediate\read5to\toto%
		\immediate\closein5%
		\immediate\openin6=Biblio\relax%
		\global\@tt=0\relax%
		\global\@ff=0\relax%
		\endtrue%
		\loop\ifeof6{}\else\immediate\read6to\repere\relax%
		\immediate\read6to\author\relax%
		\immediate\read6to\title\relax%
		\immediate\read6to\ref\relax%
		\ifx\toto\repere\relax\global\@tt=1\relax%
		\fi%
		\fi%
		\ifeof6\global\@ff=1\relax\fi%
		\ifnum\@tt=0\relax\ifnum\@ff=0\relax%
		\endtrue\else\endfalse\fi\else\endfalse\fi%
		\ifend\repeat%
		\immediate\closein6%
		}%

\def\citerr#1{\ifnum\c@biblio>0\relax\findsubr{#1}%
		\ifnum\@t=0\relax\findprinc{#1}%
		\ifnum\@tt=0\relax\immediate\write4{citation%
		"\toto" non trouvee dans la liste}%
		\immediate\write9{10000}%
		\else%
		\bibitem\repere\author\title\ref%
		\findsubr{#1}%
		\fi\fi\fi%
	\ifnum\c@biblio=0\relax%
		\findprinc{#1}%
		\ifnum\@tt=0\relax\immediate\write4{citation%
		"\toto" non trouvee dans la liste}%
		\immediate\write9{10000}%
		\else%
		\bibitem\repere\author\title\ref%
		\findsubr{#1}%
		\fi%
	\fi}%

\def\citer#1{\ifnum\c@biblio>0\relax\findsub{#1}%
		\ifnum\@t=0\relax\findprinc{#1}%
		\ifnum\@tt=0\relax\immediate\write4{citation%
		"\toto" non trouvee dans la liste}%
		10000%
		\else%
		\bibitem\repere\author\title\ref%
		\findsub{#1}%
		\fi\fi\fi%
	\ifnum\c@biblio=0\relax%
		\findprinc{#1}%
		\ifnum\@tt=0\relax\immediate\write4{citation%
		"\toto" non trouvee dans la liste}%
		10000%
		\else%
		\bibitem\repere\author\title\ref%
		\findsub{#1}%
		\fi%
	\fi}%

\def\cite#1{$[\citer{#1}]$}%

\def\Loop#1\Repeat{\def\Body{#1}\Iterate}%
\def\Iterate{\Body \let\Next=\Iterate%
	\else \let\Next=\relax\fi \Next}%

\def\sort{\immediate\openin9=sortentry%
	\immediate\openout10=sortfile%
	\c@tot=0%
	\loop\ifeof9{}\else%
	\immediate\read9to\titi%
	\global\advance\c@tot by 1\relax%
	\fi%
	\ifeof9\@f=1\relax\fi%
	\ifnum\@f=0\relax\repeat%
	\immediate\closein9%
	\global\advance\c@tot by -1\relax%
	\c@count=0%
	\c@min=0%
	\Loop\global\advance\c@count by 1\relax%
	\c@Min=10000%
	\c@loop=0%
	\immediate\openin9=sortentry%
	\loop\global\advance\c@loop by 1\relax%
	\immediate\read9to\titi%
	\ifnum\titi<\the\c@Min\relax%
		\ifnum\titi>\the\c@min\relax%
		\c@Min=\titi\relax%
		\fi%
	\fi%
	\ifnum\c@loop<\c@tot\relax\repeat%
	\c@min=\the\c@Min\relax%
	\immediate\write10{\the\c@Min}%
	\immediate\closein9%
	\ifnum\c@count<\c@tot\relax\Repeat%
	\immediate\closein9%
	\immediate\closeout10%
	}%

\catcode`\,=12
\def\voir(#1){\begingroup\global\numvoir=0\relax%
	\immediate\openout9=sortentry%
        \xvoir#1,xxx,%
	\immediate\closeout9%
	\sort%
	\immediate\openin9=sortfile%
	\c@loop=0%
	$\,[$%
	\loop\global\advance\c@loop by 1\relax%
	\immediate\read9to\toto%
	\ifnum\c@loop=1\relax%
		$\toto\,$\else%
		,$\toto\,$\fi%
	\ifnum\c@loop<\c@tot\relax\repeat%
	$]\,$%
	\immediate\closein9%
	}%

\def\fin{xxx}%
\def\xvoir#1,{\def\tempor{#1}%
        \ifx\tempor\fin\relax\endgroup%
	\else%
        \citerr\tempor%
        \expandafter\xvoir\fi%
	}%

\def\Date{\space\the\day\ \ifcase\month\or janvier\or
f\'evrier\or mars\or avril%
\or mai\or juin\or juillet%
\or ao\^ut\or septembre\or octobre%
\or novembre\or d\'ecembre}

\def\eqalign#1{\null\vcenter{\openup\jot\m@th
                \ialign{\strut\hfil$\displaystyle{##}$&
                $\displaystyle{{}##}$\hfil\crcr #1\crcr }}}

\def\xfigure#1{\midinsert%
		\global\advance\cfigure by 1\relax%
		\vbox{\centerline{\epsfbox{#1}}%
		\centerline{{\bf Figure\hskip 2mm%
		\ifx\typepartie\ann{A\the\cfigure
		\immediate\write4{Figure A\the\cfigure}
 		}\else{}\fi%
		\ifx\typepartie\chap{\the\cfigure
		\immediate\write4{Figure \the\cfigure}
		}\else{}\fi%
		}\hskip 5mm}%
		}%
		\endinsert}%

\def\caption#1{\global\advance\cfigure by 1\relax%
		{\bf Figure\hskip 2mm%
		\ifx\typepartie\ann{A\the\cfigure
		\immediate\write4{Figure A\the\cfigure}
 		}\else{}\fi%
		\ifx\typepartie\chap{\the\cfigure
		\immediate\write4{Figure \the\cfigure}
		}\else{}\fi%
		}\hskip 5mm #1
		\vskip 3mm}%

\def\figure#1#2{\ifx\typepartie\chap\relax
		\immediate\write1{\def\noexpand\typepartie{chapitre}}
		\immediate\write11{\def\noexpand%
		\typepartie%
		{chapitre}}\fi%
		\ifx\typepartie\ann\relax
		\immediate\write1{\def\noexpand\typepartie{annexe}}
		\immediate\write11{\def\noexpand%
		\typepartie%
		{annexe}}\fi%
		\immediate\write1{\noexpand\caption{#2}\relax}%
		\immediate\write11{\noexpand\xfigure{#1}\relax}}

\def\no{\global\advance\cno by 1\relax%
		\eqno%
		\ifx\typepartie\ann{(A\the\cno)
		}\else{}\fi%
		\ifx\typepartie\chap{(\the\cno)
		}\else{}\fi%
		}%

\def\zero{\cno=0\relax
	\cfootnot=0\relax
	\cfigure=0\relax
	}

\c@chapter=0\relax
\c@annexe=0\relax
\cfigure=0\relax
\cno=0\relax
\cfootnot=0\relax

\def\btext{\pageno=1\relax
		}%
\def\etext{}

\def\nopage{\headline{\hfil}%
        \footline{\hfil}}%
\def\simplepage{\headline{\hfil}%
	\footline{\hfill -\ \folio\ -\hfill}%
	}
\parindent=5mm

\magnification 1150

\beginbiblio

\bibmodif{alme}{A.P. de Almeida, F.T. Brandt, J. Frenkel}{Thermal matter
and radiation in a gravitational field}{Phys. Rev. D {\bf 49}, 4196
(\oldstyle{1994}).}

\bibmodif{alta}{M.B. Altaie, J.S. Dowker}{Spinor fields in Einstein universe:
finite temperature effects}{Phys. Rev. D {\bf 18}, 3557 (\oldstyle{1978}).}

\bibmodif{alt89}{T. Altherr, P. Aurenche, T. Becherrawy}{On infrared
and mass singularities of perturbative QCD in a quark-gluon plasma}%
{Nucl. Phys. B {\bf 315}, 436 (\oldstyle{1989}).}

\bibmodif{alt89a}{T. Altherr, P. Aurenche}{Finite temperature QCD
corrections to lepton-pair formation in a quark-gluon plasma}{Z. Phys.
C {\bf 45}, 99 (\oldstyle{1989}).}

\bibmodif{alt89b}{T. Altherr, P. Aurenche}{Fermion self-energy corrections
in perturbative theory at finite temperature}{Phys. Rev. D {\bf 40}, 4171
(\oldstyle{1989}).}

\bibmodif{alt-th}{T. Altherr}{Etude perturbative en th\'eorie quantique
des champs \`a temp\'erature finie, application \`a l'\'emission de
paires de leptons par un plasma de quarks et de gluons}{Th\`ese de doctorat.}

\bibmodif{alt2}{T. Altherr}{Infrared problem in $g\phi^4$ theory
at finite temperature}{Phys. Lett. B {\bf 238}, 360 (\oldstyle{1990}).}

\bibmodif{alt90}{T. Altherr}{Axion emission from a hot plasma}{Z. Phys.
C {\bf 47}, 559 (\oldstyle{1990}).}

\bibmodif{alt90a}{T. Altherr, T. Becherrawy}{Cancellation of infrared
and mass singularities in the thermal dilepton rate}{Nucl. Phys.
B {\bf 330}, 174 (\oldstyle{1990}).}

\bibmodif{alt91a}{T. Altherr, K. Kainulainen}{Electromagnetic interactions
and chirality flip of neutrinos in a thermal background}{Phys. Lett. B
{\bf 262}, 79 (\oldstyle{1991}).}

\bibmodif{alt91b}{T. Altherr}{Infrared singularities cancellation in reaction
rates at finite temperature}{Phys. Lett. B {\bf 262}, 314 (\oldstyle{1991}).}

\bibmodif{alt91}{T. Altherr, T. Grandou, R.D. Pisarski}{Thermal
instability in $(\phi^3)_6$}{Phys. Lett. B {\bf 271}, 183 (\oldstyle{1991}).}

\bibmodif{alt92}{T. Altherr, P.V. Ruuskanen}{Low-mass dileptons
at high momenta in ultra-relativistic heavy-ion collisions}{Nucl.
Phys. B {\bf 380}, 377 (\oldstyle{1992}).}

\bibmodif{alt93}{T. Altherr}{Introduction to thermal field theory}{Int.
Journal of Modern Physics A {\bf 8}, 5605 (\oldstyle{1993}).}

\bibmodif{alt93a}{T. Altherr, T. Grandou}{Thermal field theory and infinite
statistics}{Nucl. Phys. B {\bf 402}, 195 (\oldstyle{1993}).}

\bibmodif{alt94a}{T. Altherr}{Problems of perturbation series
in non-equilibrium quantum field theories}{Preprint cern-th.7271/94.}

\bibmodif{alt94b}{T. Altherr}{Resummation of perturbation series
in non-equilibrium scalar field theory}{Preprint cern-th.7336/94.}

\bibmodif{andr}{F. Andrews}{On the general theory of the approach to
equilibrium. III. Inhomogeneous systems}{J. of Math. Phys. {\bf 2}, 91
(\oldstyle{1961}).}

\bibmodif{aur}{P. Aurenche, T. Becherrawy}{A comparison of the real-time
and imaginary-time formalisms of finite-temperature field theory
for 2,3 and 4-point green functions}{Nucl. Phys. B {\bf 379}, 259
(\oldstyle{1992}).}

\bibmodif{aur1}{P. Aurenche, T. Becherrawy, E. Petitgirard}{Retarded/Advanced
correlation functions and soft photon production in the hard loop
approximation}{Preprint ENSLAPP-A-452/93.}

\bibmodif{baier2}{R. Baier, B. Pire, D. Schiff}{Dilepton production
at finite temperature: Perturbative treatment at order $\alpha_s$}{Phys.
Rev. D {\bf 38}, 2814 (\oldstyle{1988}).}

\bibmodif{baier}{R. Baier, G. Kunstatter, D. Schiff}{High-temperature fermion
propagator: Resummation and gauge dependance of the damping rate}{Phys. Rev. D
{\bf 45}, 4381 (\oldstyle{1992}).}

\bibmodif{baier3}{R. Baier, A. Niegawa}{Analytic continuation
of thermal N-point functions from imaginary to real energies}{Phys. Rev. D
{\bf 49}, 4107 (\oldstyle{1994}).}

\bibmodif{baier1}{R. Baier, R. Kobes}{Damping rate of a fast fermion in hot
QED%
}{Preprint ENSLAPP-A-465/94.}

\bibmodif{baner}{N. Banerjee, S. Mallik}{Spinor field theory
at finite temperature in the early universe}{Phys. Rev. D {\bf 45},
701 (\oldstyle{1992}).}

\bibmodif{beck}{R. Beckmann, F. Karsch}{Bose-Einstein condensation
of a relativistic gas in $d$ dimensions}{Phys. Rev. Lett. {\bf 43}, 1277
(\oldstyle{1979}).}

\bibmodif{boul}{D.G. Boulware}{Hawking radiation and thin shells}{Phys.
Rev. D {\bf 13}, 2169 (\oldstyle{1976}).}

\bibmodif{pis2}{E. Braaten, R.D. Pisarski}{Soft amplitudes in hot
gauge theories: a general analysis}{Nucl. Phys. B {\bf 337}, 569 (
\oldstyle{1990}).}

\bibmodif{pis3}{E. Braaten, R.D. Pisarski}{Deducing hard thermal loops
from Ward identities}{Nucl. Phys. B {\bf 339}, 310 (\oldstyle{1990}).}

\bibmodif{braat}{E. Braaten, R.D. Pisarski, T.C. Yuan}{Production of hot
dileptons in the quark-gluon plasma}{Phys. Rev. Lett. {\bf 64}, 2242
(\oldstyle{1990}).}

\bibmodif{pis4}{E. Braaten, R.D. Pisarski}{Simple effective lagrangian for
hard thermal loops}{Phys. Rev. D {\bf 45}, 1827 (\oldstyle{1992}).}

\bibmodif{branden}{R.H. Brandenberger}{Quantum field theory methods and
inflationary
universe models}{Rev. of Mod. Phys. {\bf 57}, 1 (\oldstyle{1985}).}

\bibmodif{calz2}{E. Calzetta, B.L. Hu}{Closed time path functional formalism
in curved spacetime: application to cosmological back reaction problems}{
Phys. Rev. D {\bf 35}, 495 (\oldstyle{1987}).}

\bibmodif{calz}{E. Calzetta, B.L. Hu}{Nonequilibrium quantum fields:
Closed-time-path effective action, Wigner function, and Boltzmann equation}{
Phys. Rev. D {\bf 37}, 2878 (\oldstyle{1988}).}

\bibmodif{calz1}{E. Calzetta, S. Habib, B.L. Hu}{Quantum kinetic field theory
in curved spacetime: covariant Wigner function and Liouville-Vlasov
equations}{Phys. Rev. D {\bf 37}, 2901 (\oldstyle{1988}).}

\bibmodif{cand}{P. Candelas, D.J. Raine}{Feynman propagator in curved
spacetime}{Phys. Rev. D {\bf 15}, 1494 (\oldstyle{1977}).}

\bibmodif{cani}{V. Canivel, P. Seglar}{Dynamics of the Wigner distribution
functions:
conservation of positivity in time}{Physica A {\bf 94}, 254 (\oldstyle{1978}).}

\bibmodif{chai}{M. Chaichian}{Quantum field theory at finite energy}{Nucl.
Phys. B {\bf 396}, 737 (\oldstyle{1993}).}

\bibmodif{chou}{K. Chou, Z. Su, B. Hao, L. Yu}{Equilibrium and
nonequilibrium formalisms made unified}{Phys. Rep. {\bf 118}, 1
(\oldstyle{1985}).}

\bibmodif{chu}{H. Chu, H. Umezawa}{A unified formalism of thermal
quantum field theory}{Non publi\'e.}

\bibmodif{chu1}{H. Chu, H. Umezawa}{Feynman diagram recipes in TFD}{Non
publi\'e.}

\bibmodif{chu2}{H. Chu, H. Umezawa}{Time ordering theorem and
calculational recipes for thermo field dynamics}{Phys. Lett. A {\bf 177},
385 (\oldstyle{1993}).}

\bibmodif{cley}{J. Cleymans, R.V. Gavai, E. Suhonen}{Quarks and
gluons at high temperatures and densities}{Phys. Rep. {\bf 130}, 217
(\oldstyle{1986}).}

\bibmodif{col}{S. Coleman, E. Weinberg}{Radiative corrections as the
origin of spontaneous symmetry breaking}{Phys. Rev. D {\bf 7}, 1888
(\oldstyle{1973}).}

\bibmodif{craig}{R.A. Craig}{Perturbation expansion for real-time Green's
functions}{J. of Math. Phys. {\bf 9}, 605 (\oldstyle{1968}).}

\bibmodif{dani}{P. Danielewicz}{Quantum theory of nonequilibrium processes,
I}{Ann.
of Phys. {\bf 152}, 239 (\oldstyle{1984}).}

\bibmodif{dani1}{P. Danielewicz}{Quantum theory of nonequilibrium processes,
II.
Application to nuclear collisions}{Ann. of Phys. {\bf 152}, 305
(\oldstyle{1984}).}

\bibmodif{dash}{R.F. Dashen, B. Hasslacher, A. Neveu}{Semiclassical
bound states in an asymptotically free theory}{Phys. Rev. D {\bf 12},
2443 (\oldstyle{1975}).}

\bibmodif{de-boer}{W.P.H. De Boer, Ch.G. Van Weert}{A note on Keldysh's
perturbation
formalism}{Physica A {\bf 98}, 579 (\oldstyle{1979}).}

\bibmodif{dolan}{L.Dolan, R.Jackiw}{Symmetry behavior at finite
temperature}{Phys. Rev. D {\bf 9}, 3320 (\oldstyle{1974}).}

\bibmodif{dowk}{J.S. Dowker, R. Critchley}{Vacuum stress tensor in an
Einstein universe: finite temperature effects}{Phys. Rev. D {\bf 15},
1484 (\oldstyle{1977}).}

\bibmodif{eij}{M.A. van Eijck, R. Kobes, Ch.G. van Weert}{Transformations
of real-time finite-temperature Feynman rules}{Preprint ENSLAPP-A-471/94.}

\bibmodif{elmf}{P. Elmfors, K. Enqvist, I. Vilja}{On the
non-equilibrium early universe}{Preprint NORDITA-93/80.}

\bibmodif{elz1}{H.Th. Elze, U. Heinz, K. Kjantie, T. Toimela}{High
temperature gluon matter in the background gauge}{Z. Phys. C {\bf 37}, 305
(\oldstyle{1988}).}

\bibmodif{elz}{H.Th. Elze, K. Kajantie, T. Toimela}{Chromomagnetic screening
at high temperature}{Z. Phys. C {\bf 37}, 601 (\oldstyle{1988}).}

\bibmodif{ev3}{T.S. Evans}{Zero momentum calculations for finite
temperature field theory}{Z. Phys. C {\bf 36}, 153 (\oldstyle{1987}).}

\bibmodif{ev4}{T.S. Evans}{Zero energy and thermodynamic equilibrium}{Z.
Phys. C {\bf 41},333 (\oldstyle{1988}).}

\bibmodif{ev2}{T.S. Evans}{A new time contour for equilibrium
real-time thermal field theories}{Phys. Rev. D {\bf 47}, 4196
(\oldstyle{1993}).}

\bibmodif{ev1}{T.S. Evans}{A new time contour for thermal field theories}
{Talk given at the "3rd Workshop on Thermal Field Theories and their
Applications", Banff, Canada (\oldstyle{1993}).}

\bibmodif{ev6}{T.S. Evans, I. Hardman, H. Umezawa, Y. Yamanaka}{A
time-dependent
nonequilibrium calculational sheme towards the study of temperature
fluctuations}{F. Phys. {\bf 41}, 151 (\oldstyle{1993}).}

\bibmodif{ev5}{T.S. Evans, A.C. Pearson}{A re-examination of the
path ordered approach to real time thermal field theory}{Preprint
Imperial/TP/93-94/09.}

\bibmodif{frenk}{J. Frenkel, J.C. Taylor}{High-temperature limit of
thermal QCD}{Nucl. Phys. B {\bf 334}, 199 (\oldstyle{1990}).}

\bibmodif{fuji2}{Y. Fujimoto, H. Matsumoto, H. Umezawa, I. Ojima}{Mass-%
derivative
formula and the singularity structure in thermo field dynamics}{Phys.
Rev. D {\bf 30}, 1400 (\oldstyle{1984}).}

\bibmodif{fuji}{Y. Fujimoto, R. Grigjanis}{Higher order calculation
in thermo field theory}{Z. Phys. C {\bf 28}, 395 (\oldstyle{1985}).}

\bibmodif{fuji3}{Y. Fujimoto}{A comment on the perturbation in thermo field
dynamics}{Z. Phys. C {\bf 30}, 99 (\oldstyle{1985}).}

\bibmodif{fuji1}{Y. Fujimoto}{Imaginary part in thermo field dynamics}{Phys.
Rev. D {\bf 33}, 590 (\oldstyle{1986}).}

\bibmodif{fuji4}{Y. Fujimoto, H. Yamada}{A supplementary remark on finite
temperature perturbation}{Z. Phys. C {\bf 37}, 265 (\oldstyle{1988}).}

\bibmodif{fujita1}{S. Fujita}{Equivalence between the two generalized
master equations}{Physica {\bf 28}, 281 (\oldstyle{1962}).}

\bibmodif{fujita2}{S. Fujita}{Thermodynamic evolution equation for
a quantum statistical gas}{J. of Math. Phys. {\bf 6}, 1877 (\oldstyle{1965}).}

\bibmodif{fujita}{S. Fujita}{Resolution of the hierarchy of Green's functions
for fermions}{Phys. Rev. A {\bf 4}, 1114 (\oldstyle{1971}).}

\bibmodif{gabel}{Y. Gabellini, T. Grandou, D. Poizat}{Electron-positron
annihilation in thermal QCD}{Ann. of Phys. {\bf 202}, 436 (\oldstyle{1990}).}

\bibmodif{grand1}{T. Grandou, M. Le Bellac, J.L. Meunier}{Mass
singularities at finite temperature in a scalar field theory}{Z. Phys. C
{\bf 43}, 575 (\oldstyle{1989}).}

\bibmodif{grand2}{T. Grandou, M. Le Bellac, D. Poizat}{Remarks on
infrared singularities of relativistic thermal field theories}{Phys.
Lett. B {\bf 249}, 478 (\oldstyle{1990}).}

\bibmodif{grand}{T. Grandou, M. Le Bellac, D. Poizat}{Cancellation
of infrared and collinear singularities in relativistic thermal field
theories}{Nucl. Phys. B {\bf 358}, 408 (\oldstyle{1991}).}

\bibmodif{gross}{D.J. Gross, R.D. Pisarski, L.G. Yaffe}{QCD and
instantons at finite temperature}{Rev. of Mod. Phys. {\bf 53}, 43 (%
\oldstyle{1981}).}

\bibmodif{guer}{F. Guerin}{Rules for diagrams in thermal field theories}{Phys.
Rev. D {\bf 49}, 4182 (\oldstyle{1994}).}

\bibmodif{gui1}{Y.X. Gui}{Quantum field in $\eta$-$\xi$ spacetime}{Phys.
Rev. D {\bf 42}, 1988 (\oldstyle{1990}).}

\bibmodif{gui3}{Y.X. Gui}{Fermion fields in $\eta$-$\xi$ spacetime}{Phys.
Rev. D {\bf 45}, 697 (\oldstyle{1992}).}

\bibmodif{gui2}{Y.X. Gui}{$\eta$-$\xi$ spacetime and thermo fields}{Phys.
Rev. D {\bf 46}, 1869 (\oldstyle{1992}).}

\bibmodif{gupta}{U. Gupta, A.K. Rajagopal}{Density functional formalism
at finite temperatures with some applications}{Phys. Rep. {\bf 87}, 259
(\oldstyle{1982}).}

\bibmodif{hab}{H.E. Haber, H.A. Weldon}{Thermodynamics of an
ultrarelativistic ideal bose gas}{Phys. Rev. Lett. {\bf 46}, 1497
(\oldstyle{1981}).}

\bibmodif{hab1}{H.E. Haber, H.A. Weldon}{On the relativistic
Bose-Einstein integrals}{J. Math. Phys. {\bf 23}, 1852 (\oldstyle{1982}).}

\bibmodif{hall1}{A.G. Hall}{Time-path method for the evolution of
reduced density matrices in nonequilibrium systems}{Physica A {\bf 80}, 369
(\oldstyle{1975}.}

\bibmodif{hall}{A.G. Hall}{Non-equilibrium Green functions: generalized
Wick's theorem and diagrammatic perturbation theory with initial
correlations}{J. Phys. A {\bf 8}, 214 (\oldstyle{1975}).}

\bibmodif{hart}{J.B. Hartle, S.W. Hawking}{Path-integral derivation
of black-hole radiance}{Phys. Rev. D {\bf 13}, 2188 (\oldstyle{1976}).}

\bibmodif{haw1}{S.W. Hawking}{Particle creation by black holes}{Comm.
Math. Phys. {\bf 43}, 199 (\oldstyle{1975}).}

\bibmodif{haw2}{S.W. Hawking}{The development of irregularities in a single
bubble inflationary universe}{Phys. Lett. B {\bf 115}, 295 (\oldstyle{1982}).}

\bibmodif{henin}{F. Henin, P. Resibois, F. Andrews}{On the general theory of
the
approach to equilibrium. II. Interacting particles}{J. of Math. Phys. {\bf 2},
68
(\oldstyle{1961}).}

\bibmodif{hu}{B.L. Hu, R. Critchley, A. Stylianopoulos}{Finite
temperature field theory in curved spacetime: Quasilocal
effective Lagrangians}{Phys. Rev. D {\bf 35}, 510 (\oldstyle{1987}).}

\bibmodif{itz}{C. Itzykson, J.B. Zuber}{Quantum field theory}{Ed.
Mac Graw Hill}

\bibmodif{kan}{T. Kaneko}{Zero momentum limits of two loop finite
temperature self-energies in $\phi^4$ and $\phi^3$ coupling theories}{Phys.
Rev. D {\bf 49}, 4209 (\oldstyle{1994}).}

\bibmodif{keil}{W. Keil}{Radiative corrections and renormalization
at finite temperature: a real time approach}{Phys. Rev. D {\bf 40},
1176 (\oldstyle{1989}).}

\bibmodif{keld}{L.V. Keldysh}{Diagram technique for
nonequilibrium processes}{Sov. Phys. JETP {\bf 20}, 1018 (\oldstyle{1964}).}

\bibmodif{kobes}{R.L. Kobes}{Correspondance between imaginary-time
and real-time finite-temperature field theory}{Phys. Rev. D {\bf 42},
562 (\oldstyle{1990}).}

\bibmodif{kobes1}{R.L. Kobes}{Three-point function at finite temperature
in the real time formalism}{Phys. Rev. Lett. {\bf 67}, 1384 (\oldstyle{1991}).}

\bibmodif{kk}{R.L. Kobes, G. Kunstatter, A. Rebhan}{QCD plasma parameters
and the gauge-dependant gluon propagator}{Phys. Rev. Lett. {\bf 64}, 2992
(\oldstyle{1990}).}

\bibmodif{kb1}{R.L. Kobes, G.W. Semenov}{Discontinuities of green functions
in field theory at finite temperature and density}{Nucl. Phys. B {\bf 260},
714 (\oldstyle{1985}).}

\bibmodif{kob-kow}{R.L. Kobes, K.L. Kowalski}{Path-integral formulation
of real-time finite-temperature field theory}{Phys. Rev. D {\bf 34},
513 (\oldstyle{1986}).}

\bibmodif{kb2}{R.L. Kobes, G.W. Semenov}{Discontinuities of green functions
in field theory at finite temperature and density, II}{Nucl. Phys. B {\bf 272},
329 (\oldstyle{1986}).}

\bibmodif{kob-sem}{R.L. Kobes, G.W. Semenov, N. Weiss}{Real-time
Feynman rules for gauge theories with fermions at finite temperature
and density}{Z. Phys. C {\bf 29}, 371 (\oldstyle{1985}).}

\bibmodif{koren}{V. Korenman}{On perturbation expansions for real-time
Green's functions}{J. of Math. Phys. {\bf 10}, 1387 (\oldstyle{1969}).}

\bibmodif{krae}{U. Kraemmer, A.K. Rebhan, H. Schulz}{Resummations
in hot scalar electrodynamics}{Preprint ITP-UH-01/94.}

\bibmodif{kukha}{Y.A. Kukharenko, S.G. Tikhodeev}{Diagram technique
in the theory of relaxation processes}{Sov. Phys. JETP {\bf 56}, 831
(\oldstyle{1982}).}

\bibmodif{land5}{L.D. Landau, E.M. Lifchitz}{Physique statistique, partie I}{
Cours de physique th\'eorique, tome 5, Ed. MIR}

\bibmodif{land9}{L.D. Landau, E.M. Lifchitz}{Physique statistique, partie II}{
Cours de physique th\'eorique, tome 9, Ed. MIR}

\bibmodif{land10}{L.D. Landau, E.M. Lifchitz}{Cin\'etique physique}{Cours de
physique th\'eorique, tome 10, Ed. MIR}

\bibmodif{Landsh}{P.V. Landshoff, J.C. Taylor}{Photon radiation in a heat
bath}{Preprint DAMTP 94/2.}

\bibmodif{lands1}{N.P. Landsman}{Consistent real-time propagators for
any spin, mass, temperature and density}{Phys. Lett. B {\bf 172}, 46
(\oldstyle{1986}).}

\bibmodif{lands}{N.P. Landsman, Ch.G. van Weert}{Real and imaginary
time field theory at finite temperature and density}{Phys. Rep.
{\bf 145}, 141 (\oldstyle{1987}).}

\bibmodif{lands2}{N.P. Landsman}{Hilbert space and propagator in thermal
field theory}{Phys. Rev. Lett. {\bf 60}, 1909 (\oldstyle{1988}).}

\bibmodif{lebell1}{M. Le Bellac, D. Poizat}{Renormalization of external lines
in relativistic field theories at finite temperature}{Z. Phys. C {\bf 47},
125 (\oldstyle{1990}).}

\bibmodif{lebell}{M. Le Bellac, P. Reynaud}{Cancellation of infrared
and collinear singularities in relativistic thermal field theories (II)}{Nucl.
Phys. B {\bf 380}, 423 (\oldstyle{1992}).}

\bibmodif{lee1}{L.Y. Lee}{Evaluation of the one-loop effective
action at zero and finite temperature: Scalar fields}{Phys. Rev. D {\bf 49},
4101 (\oldstyle{1994}).}

\bibmodif{lee}{T.D. Lee, M. Nauenberg}{Degenerate systems and
mass singularities}{Phys. Rev. {\bf 133}, 1549 (\oldstyle{1964}).}

\bibmodif{lim}{S.C. Lim}{Finite temperature field from stochastic
mechanics}{Phys. Lett. B {\bf 188}, 239 (\oldstyle{1987}).}

\bibmodif{linde}{A.D. Linde}{Infrared problem in the thermodynamics
of the Yang-Mills gas}{Phys. Lett. B {\bf 96}, 289 (\oldstyle{1980}).}

\bibmodif{makh}{A. Makhlin}{Nonequilibrium quantum field kinetics}{Preprint
SUNY-NTG-92-11.}

\bibmodif{mari}{M. Marinaro}{Perturbative expansions at finite
temperature}{Phys. Rep. {\bf 137}, 81 (\oldstyle{1986}).}

\bibmodif{matsu}{H. Matsumoto,
I. Ojima, H. Umezawa}{Perturbation and renormalisation in thermo
field dynamics}{Ann. Phys. {\bf 152}, 348 (\oldstyle{1984}).}

\bibmodif{matsu1}{H. Matsumoto, Y. Nakano, H. Umezawa}{Free energy in
thermo field dynamics}{Phys. Rev. D {\bf 31}, 1495 (\oldstyle{1985}).}

\bibmodif{nadk}{S. Nadkarni}{Non-Abelian Debye screening: the color
averaged potential}{Phys. Rev. D {\bf 33}, 3738 (\oldstyle{1986}).}

\bibmodif{nieg}{A. Niegawa}{Path-integral formulation of real-time
quantum field theories at finite temperature}{Phys. Rev. D {\bf 40}, 1199
(\oldstyle{1989}).}

\bibmodif{nieg1}{A. Niegawa, K. Takashiba}{Cancellation of infrared
singularities in finite-temperature reaction rates}{Nucl. Phys. B {\bf 370},
335 (\oldstyle{1992}).}

\bibmodif{niem1}{A.J. Niemi, G.W. Semenov}{Thermodynamic calculations
in relativistic finite-temperature field theories}{Nucl. Phys. B {\bf 230},
181 (\oldstyle{1984}).}

\bibmodif{niem}{A.J. Niemi}{Nonequilibrium quantum field theories}{Phys.
Lett. B {\bf 203}, 425 (\oldstyle{1987}).}

\bibmodif{papp}{E. Papp}{Quasiclassical approach to the
virial theorem and to the evaluation of the ground state energy}{Phys.
Rep. {\bf 136}, 103 (\oldstyle{1986}).}

\bibmodif{parw}{R.R. Parwani}{Resummation in hot scalar field
theory}{Phys. Rev. D {\bf 45}, 4695 (\oldstyle{1992}).}

\bibmodif{pis6}{R.D. Pisarski}{Computing finite-temperature loops
with ease}{Nucl. Phys. B {\bf 309}, 476 (\oldstyle{1988}).}

\bibmodif{pis1}{R.D. Pisarski}{Scattering amplitudes in hot gauge
theories}{Phys. Rev. Let. {\bf 63}, 1129 (\oldstyle{1989}).}

\bibmodif{pis5}{R.D. Pisarski}{Resummation of the gluon damping rate
in hot QCD}{Talk presented at Quark Matter \oldstyle{1990}.}

\bibmodif{plos}{M. Ploszajczak, M.J. Rhoades-Brown}{Nonequilibrium
truncation sheme for the statistical mechanics of relativistic matter}{%
Phys. Rev. D {\bf 33}, 3686 (\oldstyle{1986}).}

\bibmodif{poiz}{D. Poizat}{Singularit\'es infrarouges et singularit\'es de
masse dans les th\'eories quantiques des champs relativistes \`a
temp\'erature finie}{Th\`ese de doctorat.}

\bibmodif{prigo}{I. Prigogine, R. Balescu}{Irreversible processes in gases,
I. The diagram technique}{Physica {\bf 25}, 281 (\oldstyle{1959}).}

\bibmodif{prigo1}{I. Prigogine, R. Balescu}{Irreversible processes in gases,
II. The equations of evolution}{Physica {\bf 25}, 302 (\oldstyle{1959}).}

\bibmodif{prigo2}{I. Prigogine, F. Henin}{On the general theory of the
approach to equilibrium. I. Interacting normal modes}{J. of Math. Phys. {\bf
1}, 349
(\oldstyle{1960}).}

\bibmodif{prigo3}{I. Prigogine, P. Resibois}{On the kinetics of the approach
to equilibrium}{Physica {\bf 27}, 629 (\oldstyle{1961}).}

\bibmodif{ram}{J. Rammer, H. Smith}{Quantum field-theoretical methods
in transport theory of metals}{Rev. of Modern Physics {\bf 58}, 323
(\oldstyle{1986}).}

\bibmodif{ramo}{P. Ramond}{Field theory: a modern primer}{Ed. Addison
Wesley.}

\bibmodif{rebh}{A. Rebhan}{Comment on "High temperature fermion propagator:
Resummation and gauge dependance of the damping rate"}{Phys. Rev. D {\bf 46},
4779 (\oldstyle{1992}).}

\bibmodif{resib}{P. Resibois}{Approach to equilibrium in quantum systems}{Phys.
Rev. Lett. {\bf 5}, 411 (\oldstyle{1960}).}

\bibmodif{resib1}{P. Resibois}{On the equivalence between two
generalized master equations}{Physica {\bf 29}, 721 (\oldstyle{1963}).}

\bibmodif{resib2}{P. Resibois, M. Mareschal}{Kinetic equations, initial
conditions and
time-reversal: a solvable one-dimensional model revisited}{Physica A {\bf 94},
211
(\oldstyle{1978}).}

\bibmodif{sem}{G.W. Semenoff, H. Umezawa}{Functional methods in
thermofield dynamics: a real time perturbation theory for quantum
statistical mechanics}{Nucl. Phys. B {\bf 220}, 196 (\oldstyle{1983}).}

\bibmodif{svet}{B. Svetitsky}%
{Symmetry aspects of finite temperature confinement transitions}%
{Phys. Rep. {\bf 132}, 1 (\oldstyle{1986}).}

\bibmodif{umez}{H. Umezawa}{Equilibrium and non-equilibrium thermal
physics}{Non
publi\'e.}

\bibmodif{unr}{W.G. Unruh}{Notes on black-hole evaporation}{Phys. Rev. D
{\bf 14}, 870 (\oldstyle{1976}).}

\bibmodif{vanhove2}{L. Van Hove}{The approach to equilibrium in quantum
statistics:
a perturbaton treatment to general order}{Physica {\bf 23}, 441
(\oldstyle{1957}).}

\bibmodif{vanhove1}{L. Van Hove}{The ergodic behaviour of quantum many-body
systems}{Physica {\bf 25}, 268 (\oldstyle{1959}).}

\bibmodif{vanhove}{L. Van Hove}{Quantum field theory at
positive temperature}{Phys. Rep. {\bf 137}, 11 (\oldstyle{1986}).}

\bibmodif{weld1}{H.A. Weldon}{Covariant calculations at finite temperature.
The relativistic plasma.}{Phys. Rev. D {\bf 26}, 1394 (\oldstyle{1982}).}

\bibmodif{weld}{H.A. Weldon}{Effective fermion masses of order gT in
hight-temperature gauge theories with exact chiral invariance}{Phys.
Rev. D {\bf 26}, 2789 (\oldstyle{1982}).}

\bibmodif{weld3}{H.A. Weldon}{Simple rules for discontinuities in
finite-temperature fiels theory}{Phys. Rev. D {\bf 28}, 2007
(\oldstyle{1983}).}

\bibmodif{weld4}{H.A. Weldon}{Bloch-Nordsiek cancellation of infrared
divergences at finite temperature}{Phys. Rev. D {\bf 44}, 3955
(\oldstyle{1991}).}

\bibmodif{weld2}{H.A. Weldon}{Thermalization of boson propagators in
finite-temperature field theory}{Phys. Rev. D {\bf 45}, 352 (\oldstyle{1992}).}

\bibmodif{weld5}{H.A. Weldon}{Suppression of bremsstrahlung at nonzero
temperature}{Phys. Rev. D {\bf 49}, 1579 (\oldstyle{1993}).}

\endbiblio
\btext
\nopage
\footline{\hfill}
\null\vfill
\centerline{\bf THE EFFECT OF THE VERTICAL PART OF THE PATH}
\centerline{\bf ON THE REAL TIME FEYNMAN RULES}
\centerline{\bf IN FINITE TEMPERATURE FIELD THEORY}
\vfill
\centerline{{\bf Fran\c cois Gelis}\footnot{
\'el\`eve \`a l'\'Ecole Normale Sup\'erieure de Lyon.
E-mail: fgelis@ens.ens-lyon.fr}}
\centerline{\it Laboratoire de Physique Theorique ENSLAPP}
\centerline{\it URA 1436 du CNRS associ\'ee \`a l'ENS de Lyon}
\vfill
\centerline{In memory of Tanguy Altherr}
\vfill
\centerline{\bf Abstract}

The effect of the contribution of the vertical part of the real time path
is studied completely in the case of two points functions. Indeed,
this vertical
part generally contributes in the calculation of a given graph. Moreover,
this contribution is essential in order to have a consistent equilibrium
theory: thanks to this contribution, the Green function are effectively
invariant by time translation, as they should be. As a by product, it is
shown that the perturbative calculations give a result which does not
depend on the initial time $t_I$ and final time $t_F$ of the path. The
property of independence with respect to $t_I$ is closely related to
the KMS conditions, i.e. to the fact the system is in thermal equilibrium.
In the case of two point functions,
the contribution of the vertical part can be taken into account by the
$n(|k_0|)$ prescription in the usual RTF Feynman rules.
\vfill
\centerline{\bf \hfill ENSLAPP-A-501/94}
\centerline{\bf \hfill hep-ph/9412347}
\vfill\eject

\pageno=1
\simplepage
\chapter{Introduction}

Quantum Field Theory can be formulated on every descending
path in complex time
plane, provided it begins at an initial time $t_I$ and ends at the time
$t_I-i\beta$.
Among these paths, of particularly important use are the so-called
 real-time paths,
represented on figure $1$.
This kind of path has two parts parallel to the real
time axis, and two vertical parts. In the following, we will  only consider
in this familly of paths the path corresponding to
 $\sigma=0$\footnot{Nevertheless, most of the calculations made in this paper
are valid whatever is the path going from $t_I$ to $t_I-i\beta$.}
(see figure $2$), so that
there will be only one vertical segment, which will be refered in the
following as the "vertical part" and denoted simply $C_v$ instead of
$C_4$.

It is necessary to find a proper way to get rid of the vertical part
if one wants to have simple Feynman rules in Fourier space.The usual way
to achieve that
\voir({kob-kow},{kob-sem},{lands},{mari},{niem1},{niem},{vanhove})
 has been to simply drop the vertical part, arguing that it does
not contribute to the Green functions in the limit $t_I\rightarrow -\infty$,
$t_F\rightarrow +\infty$, thanks to the Riemann-Lebesgue lemma which is
said to imply the factorization of the generating functional: $Z[j]=
Z_{12}[j]\,Z_{34}[j]$. However, it has been noticed that the vertical part
can contribute to the Green functions, for example in vacuum graphs
\voir({ev3},{ev4}) and even in every two point function
\cite{nieg}. Alternative paths with only two
branchs have been introduced \cite{ev2} in order
to justify  the $2\times 2$ matrix formalism, but the limiting
procedure to obtain Real Time Formalism in Fourier space (simply called
RTF in the following)
 is not completely transparent.
In section
{\uppercase\expandafter{\romannumeral 2}}, we
see by a canonical approach (i.e. without
the path integral formalism) how this vertical part does arise, and what
is the precise meaning of its suppression. We show that dropping
the vertical part is equivalent to perform averages with the free density
operator $\exp(-\beta H_0)$ instead of $\exp(-\beta H)$, the dynamics of
the fields still being governed by the complete Hamiltonian.

Another problem related to the RTF
 is the so called $n(|k_0|)$ prescription. Indeed,
the free propagators contain the statistical function $n(\cdot)$ always
accompanied with a factor $\delta(k_0^2-\omega_k^2)$, where $\omega_k=\sqrt{
{\ibf k}^2+m^2}$. So, in the free propagators, it makes no difference
to have $n(|k_0|)$ or $n(\omega_k)$.
But, because the cancellation of pathologies generates
derivatives of the $\delta(\cdot)$ distributions, the choice of the argument
of the statistical factors is very important in perturbative calculations where
pathologies arise. It has been shown that in order to preserve the KMS
relations when the propagators are regularized by a finite parameter $\epsilon
>0$, the $n(|k_0|)$ prescription is {\sl necessary} \voir({lands},{ev2}).

In \cite{nieg}, it is noticed that the vertical part contributes
to every connected
two point diagram, and that this contribution can be included among the
other terms (those which come from the horizontal parts of the path) by
simply substituing in them $n(|k_0|)$ to $n(\omega_k)$. So it seems that
the two announced problems are closely related, and that the $n(|k_0|)$
prescription is in fact the consequence of the vertical part contribution.
However, the proof given in \cite{nieg} in order to establish this property
 for the general two point diagram still uses the limit $t_I\rightarrow -
\infty$, $t_F\rightarrow +\infty$ in association with the Riemann-Lebesgue
lemma.

In section {\uppercase\expandafter{\romannumeral 3}}
 of the present paper, we perform simple
calculations on the complete real time path, with finite $t_I$ and $t_F$,
and then Fourier transform the result in order to see what is the effect
of the vertical part. We reobtain the {\sl result} announced
by \cite{nieg}. Moreover,
we also see that the result of the calculation is {\sl independent\/}
 of the times
$t_I$ and $t_F$, so that the proofs based on the limits $t_I\rightarrow
-\infty$, $t_F\rightarrow +\infty$ seems suspect. In
section {\uppercase\expandafter{\romannumeral 4}}, we
 prove explicitly that every diagram
is independent
of $t_I$ and $t_F$. The independence with respect to $t_F$ is simply
related to causality, because $t_F$ must be chosen greater than any other time
in the diagram. The independence with respect to $t_I$ is related to the fact
that the system is in equilibrium, and is proven by use of the KMS relations.

In section {\uppercase\expandafter{\romannumeral 5}}, we
prove that the vertical part
contributions can be taken into account by the $n(|k_0|)$ prescription
for every two point function, without invoking the $t_I\rightarrow -\infty$
and $t_F\rightarrow +\infty$ limits.

In section \uppercase\expandafter{\romannumeral 6}, we show that it
is possible to have a contribution of the vertical part in $n$ point
functions $(n>2)$, for certain configurations of the external momenta.

\chapter{Canonical derivation of the generating functional}

The path-integral technique, in association with the Feynman-Matthews-Salam
formula \cite{lands}, is the most powerful way to quantize a classical field.
Nevertheless, in the case of thermal field theories, a canonical approach
can clarify the actual role of the vertical branch of the path. We reproduce
 here this derivation in the simple case of a real scalar field\footnot{
Of course, the case of more complicated fields like non abelian gauge fields
is prohibitively tedious to treat by this method. But we expect that the
interpretation of the vertical part obtained here for scalar fields still
holds for other kind of fields.}\voir({dani},{lands},{ram}).

\section{The generating functional for a general average}

\def\phii{\phi_{\scriptscriptstyle I}}
 Let $\phi(x)$ be the field operator in the Heisenberg
picture, and $\phii(x)$ the field operator in the interaction picture, the
interaction
being switched on at the time $t_I$. They are related by:
$$\left\{\eqalign{&\phi(x)=U(t_I,t)\phii(x)U(t,t_I)\cr
&U(t_2,t_1)={\rm T}_c\exp i\int\limits_{t_1}^{t_2}
{\cal L}_I(\phii(x))d^4 x\cr}\right.\no$$
where the integration is performed along a path going from $t_1$ to
$t_2$ in the complex-time plane, and where ${\cal L}_I(\cdot)$ is
the interacting part of the Lagrangian. The operator $U(t_1,t_2)$
verifies the following basic properties:
$$\left\{\eqalign{&U(t_1,t_1)={\ibf 1}\cr
&U(t_1,t_2)U(t_2,t_3)=U(t_1,t_3)\cr}\right.\no$$
By introducing an arbitrary time $t_F$, we can write:
$$\phi(x)=U(t_I,t_F)U(t_F,t)\phii(x)U(t,t_I) \no$$
from which we deduce:
$${\rm T}_c\phi(x_1)\cdots\phi(x_n)={\rm T}_c\left[
\phii(x_1)\cdots\phii(x_n)\;\exp i\int\limits_{C_1\oplus C_2}^{} d^4x
\;{\cal L}_I(\phii(x))\right] \no$$
where $C_1$ is a path going along the real time axis from $t_I$ to $t_F$,
and $C_2$ the reverse path from $t_F$ to $t_I$.

Now, we define the Green functions to be the averages of the path
ordered products calculated in $(4)$, an average operator being
in general a linear form $\omega(\cdot)$, normalized to
$\omega({\ibf 1})=1$. From $(4)$, we can directly write the
generating functional of these Green functions:
$$\omega({\rm T}_c \phi(x_1)\cdots\phi(x_n))=
{1\over{Z^{\omega}[0]}}\;
{{\delta\hfil}\over
{i\delta j(x_1)}}\cdots{{\delta\hfil}\over{i\delta j(x_n)}}
Z^{\omega}[j]_{|\scriptscriptstyle j=0} \no$$
$$Z^{\omega}[j]=\omega\left({\rm T}_c\exp i \int\limits_{C_1\oplus
C_2}^{} d^4x\; {\cal L}_I(\phii(x))+j(x)\phii(x)\right) \no$$
In particular, we will denote:
\def\omh{\omega_{\scriptscriptstyle H}}
\def\omho{\omega_{\scriptscriptstyle H_0}}
\def\Tr{{\rm Tr}}
$$ \omh(\cdot)\equiv{{\Tr(e^{-\beta H}\cdot)}\over{\Tr(e^{-\beta H})}}
\qquad \omho(\cdot)\equiv{{\Tr(e^{-\beta H_0}\cdot)}\over
{\Tr(e^{-\beta H_0})}} \no $$
where $H_0$ is the free part of the Hamiltonian $H=H_0+H'$ in the
Heisenberg picture.

\section{The introduction of the vertical part}

The fundamental formula \cite{ram}, useful to perform equilibrium
 averages (\ie with
$\omh(\cdot))$, is:
$$e^{-\beta H}=e^{-\beta H_0}{\rm T}_c \exp i\int\limits_{C_v}^{} d^4x
\;{\cal L}_I(\phii(x)) \no $$
where $C_v$ is a path going from $t_I$ to $t_I-i\beta$, and which will
be taken parallel to the imaginary time axis (figure 2).
 Then, thanks to the formula
$(8)$, we can rewrite the generating functional of the equilibrium
Green functions as:
$$Z^H[j]={{\Tr(e^{-\beta H_0})}\over{\Tr(e^{-\beta H})}}\; \omho\left(
{\rm T}_c \exp i \int\limits_{C}^{} d^4x\; {\cal
L}_I(\phii(x))+j(x)\phii(x)\right) \no$$
where $C=C_1\oplus C_2\oplus C_v$ is the real time path with $\sigma=0$. If,
instead of calculating equilibrium averages, we decide to perform
$\omho(\cdot)$ averages, we do not need to use the formula $(8)$ and we
can write directly:
$$Z^{H_0}[j]= \omho\left(
{\rm T}_c \exp i \int\limits_{C_1\oplus C_2}^{} d^4x\; {\cal
L}_I(\phii(x))+j(x)\phii(x)\right) \no$$

So, apart from a $j$-independent factor which plays no role in the
perturbative calculation
of a Green function with $(5)$, we see that the only difference between
the last two formulae is the integration path. The last one shows that
when we neglect the vertical part, we are in reality calculating averages
with the free density operator $\exp(-\beta H_0)$.

We can further simplify these generating functionals by the same manipulations
as the zero temperature ones \cite{itz}, and by using the Wick's
 theorem which remains
valid for $\omho(\cdot)$ averages of free fields (see appendix $1$).
Indeed, we have:
$$\omho\left({\rm T}_c \exp i\int\limits_{\gamma}^{} d^4x\;
j(x)\phii(x)\right)=
\exp -{1\over 2}\int\limits_{\gamma}^{} d^4x\, d^4y\; j(x)j(y) \omho({\rm T}_c
\phii(x)\phii(y)) \no $$
where ${\gamma}$ is an arbitrary path, so $(11)$ applies both to $(9)$
and $(10)$. $(11)$ implies that to evaluate a graph, we need
to attach a free propagator $G_0(x,y)=\omho({\rm T}_c\phii(x)\phii(y))$ to
each line, and to integrate over the coordinates of each vertex, the
integral over time being performed along the path ${\gamma}$.

So, the important conclusion of this part is that the suppression of
 the vertical part of the time path is equivalent to perform only
$\omho(\cdot)$ averages. More precisely, we can say that adding the vertical
part allows us to do at the same time the perturbative expansion
in the dynamics of the fields (in the same way as in the zero temperature
expansion) {\sl and\/} in their statistics (\ie in the density operator,
which also contain the coupling constant):
the vertical part enables us
to have a true equilibrium theory {\sl at each order\/} of the perturbative
expansion.

\chapter{A simple example: the insertion of mass terms}

\section{Calculation in time coordinates}

\subsection{Introduction and notations}

In order to see what the effect of the vertical part in effective
 calculations is, we will do the simplest calculation which consists
in the convolution of two propagators:
$$F(x_1,x_2)=\int\limits_{C}^{}d^4x\;G_0(x_1,x)G_0(x,x_2) \no$$
This is typically the calculation one would do to insert a mass term on
the propagator of a particle. We can further simplify the computations
by using a Fourier representation with respect to the spatial
dependence:
$$G_0(t_1,t_2;{\ibf k})=\int d^3{\ibf x}\;
e^{-i{\ibf k}\cdot{\ibf x}}\;
G_0(t_1,{\ibf 0},t_2,{\ibf x})\no $$
It is well known that\footnot{Whenever a two point function
 depends on ${\ibf k}$ only through $\omega_k$, we will recall this
by replacing ${\ibf k}$ by $\omega_k$ in the list of its arguments.}:
$$
\eqalign{G_0(t_1,t_2;\omega_k)\equiv G_0(t_1,t_2;{\ibf
 k})={1\over{2\omega_k}}\Big[
&e^{-i\omega_k(t_2-t_1)}(\theta_c(t_2-t_1)+n_B(\omega_k))+
\quad\cr
&\quad
e^{-i\omega_k(t_1-t_2)}(\theta_c(t_1-t_2)+n_B(\omega_k))\Big]\cr}
 \no $$
So, the Feynman rules will be slightly modified and  we
have now to compute:
$$F(t_1,t_2;\omega_k)=\int\limits_{C}^{}dt\,G_0(t_1,t;\omega_k)
G_0(t,t_2;\omega_k) \no$$

\subsection{The $t_F-$independence}

Before going to the
calculation itself, we make a general remark on the independence
of the Green functions with respect to
 the time $t_F$. To see this property, we can choose
another final time $t'_F>t_F$. Then if $G_{t_F}$ is the Green function
calculated on a path which final time is $t_F$, we will have:
$$G_{t'_F}-G_{t_F}=\int\limits_{\gamma}^{} g(t)dt \no $$
where $\gamma$ is a path going from $t_F$ to $t'_F$ and then back to $t_F$ and
$g(t)$ a function which does not contain explicitly $t_F$ or $t'_F$. Moreover,
if $t_F$ is greater than all the other times in the problem, $g(t)$ will take
the same values on the upper and lower part of $\gamma$, and $G_{t'_F}$
will be equal to $G_{t_F}$, as announced.

\subsection{Result and commentaries}

After some straightforward calculations, we find:
$$\eqalign{F(t_1,t_2&;\omega_k)={1\over{4\omega_k^2}}\left\{\right.
n_B^2(\omega_k)\left[\right.{\beta\over i}(e^{i\omega_k(t_2-t_1)}+
e^{-i\omega_k(t_2-t_1)})\cr
&\qquad+{1\over{2i\omega_k}}(e^{i\omega_k(t_1+t_2-2t_I)}(1-e^{-2\omega_k\beta})-
e^{-i\omega_k(t_1+t_2-2t_I)}(1-e^{2\omega_k\beta}))\left.\right]\cr
&+n_B(\omega_k)\left[\right.(t_2-t_1)(e^{-i\omega_k(t_2-t_1)}-e^{i\omega_k(
t_2-t_1)})\cr
&\qquad+\left({1\over{i\omega_k}}+{\beta\over
i}\right)(e^{-i\omega_k(t_2-t_1)}+e^{i\omega_k(
t_2-t_1)})\cr
&\qquad-{1\over{i\omega_k}}(e^{i\omega_k(t_1+t_2-2t_I)}e^{-2\omega_k\beta}+
e^{-i\omega_k(t_1+t_2-2t_I)})\left.\right]\cr
&+\left[\right.\theta_c(t_2-t_1)(t_2-t_1+{1\over{i\omega_k}})e^{-i
\omega_k(t_2-t_1)}+\theta_c(t_1-t_2)(t_1-t_2+{1\over{i\omega_k}})e^{i\omega_k(
t_2-t_1)}\cr
&\qquad-{1\over{2i\omega_k}}(e^{i\omega_k(t_1+t_2-2t_I)}e^{-2\omega_k\beta}+
e^{-i\omega_k(t_1+t_2-2t_I)})\left.\right]\left.\right\}
} \no$$
Firstly, we notice that there appears some terms which are not invariant by
time translation, depending on $t_1+t_2-2t_I$. But, we verify that
the sum of all these terms is zero. Therefore, the final result is invariant
by time translation and independent of the initial time $t_I$.
We must precise that the contribution
of the vertical part\footnot{The separation between the contributions of the
horizontal part and of the vertical part has no intrinsic meaning: it depends
on the position of the external times on the path
(if they are all on the vertical part, the entire result comes from the
vertical part). So, here and in the following, the contribution of the vertical
part is defined to be
the one obtained when the external times are on the
horizontal branchs.}
(terms in
$\beta/ i$ and in $\exp(\pm 2\beta\omega_k)$) is  essential in order
to have this cancellation.
We will
generalize this property to a general graph in the next section.
Another property of this result is the fact that it remains contributions
of the vertical
part (terms proportional to $\beta/i$), and we have no
possibility to make them disappear by the limit procedure $t_I\rightarrow
-\infty$, $t_F\rightarrow +\infty$ because the result does not contain
$t_I$ or $t_F$ anymore.

We can also verify explicitly the
mass-derivative formula \voir({fuji2},{nieg}):
$$i{{\partial\hfil}\over{\partial m^2}}G_0(t_1,t_2;\omega_k)=
\int\limits_{C}^{}dt\; G_0(t_1,t;\omega_k)G_0(t,t_2;\omega_k) \no $$
It is important to notice here that this formula holds only if the
integration is performed on the {\sl complete\/} path $C$, because the
left-hand side of $(18)$ contains those terms in $\beta/ i$
exhibited above.

\section{The result in Fourier space}

Now, the following step is to perform the Fourier transformation of
$F(t_1,t_2;{\ibf k})$
with respect to the difference $t_2-t_1$\footnot{This Fourier transformation is
legitime once we have verified that the result is invariant by time
translation. Moreover, the function $F(\cdot)$ being independant of
the times $t_I$ and $t_F$, we are allowed to perform an usual Fourier
transformation with times varying from $-\infty$ to $+\infty$. We then define
a $2\times 2$ matrix according to the position of the external times on
$C_1$ or $C_2$, in the usual way.}
in order to see which  are the Feynman rules
in energy space (they are not obvious because of the contribution
of the vertical part in the result). We will denote $D_{ij,\omega}(k)$
the usual real-time matrix propagator (with $\sigma=0$)
 with distribution
 function $n_B(\omega_k)$ and $D_{ij,k_0}(k)$ the same propagators with
distribution function $n_B(|k_0|)$ (see appendix 2).
These two kinds of {\sl free} propagators are
equal because each distribution function is accompanied by a $\delta(k_0^2-
\omega_k^2)$ factor.
Then, by a simple but tedious calculation, we
 find for the Fourier transformation:
$$F_{ij}(k_0,\omega_k)=\left(D_{\omega}(k)\tau_3
 D_{\omega}(k)\right)_{ij}+{\beta\over i}
{{n_B(\omega_k)(1+n_B(\omega_k))}\over{2\omega_k}}2\pi\delta(k^2-m^2) \no $$
where the last term is the Fourier transformation of the vertical part
 contribution.
Moreover, in the first term, the cancellation of pathologies makes derivatives
of delta distributions appear (see appendix 3 and section {\uppercase%
\expandafter{\romannumeral 5}}), so that this term will not be the same if we
replace $\omega_k$ by $k_0$ in it. But, fortunately, if we do such a
 substitution, we obtain a supplementary term which cancels exactly the second
term of $(19)$, so that \cite{nieg}:
$$F_{ij}(k_0,\omega_k)=\left(D_{k_0}(k)\tau_3
 D_{k_0}(k)\right)_{ij} \no $$

In this simple example, we have seen that dropping the vertical part
contribution is e\-qui\-va\-lent to use $\omega-$type propagators
in the usual real-time Feynman rules \voir({lands},{kob-kow},{kob-sem}),
 and that we can take
this contribution into account by
using $k_0-$type propagators with the same Feynman rules.

\section{Relation with part \uppercase\expandafter{\romannumeral 2}}

To end this
section, we verify the physical interpretation of section
\uppercase\expandafter{\romannumeral 2}, \ie the fact
that dropping the vertical part is equivalent to perform averages with the
free statistics. For that purpose, we treat perturbatively an
additional mass term $-\mu^2 \phi^2 /2$ (figure 3).
If we write the free matrix
propagator as:
$$D_{\omega [k_0]}(k)=U_{\omega [k_0]}\pmatrix{
&i(k^2-m^2+i\epsilon)^{-1} & 0 \cr
&0 &-i(k^2-m^2-i\epsilon)^{-1} \cr}U_{\omega [k_0]} \no$$
where the matrix $U$ contains the distribution functions and
verifies $U\tau_3 U=\tau_3$. The only effect
of the summation of all the extra mass insertions is to replace
$m^2$ by $m^2+\mu^2$ in the central matrix. Consider for example the $11$
component of the resummed propagator:
$${\cal D}_{11,\omega[k_0]}(k)={i\over{k^2-m^2-\mu^2+i\epsilon}}+
2\pi n_B(\omega_k [|k_0|])\delta(k^2-m^2-\mu^2) \no $$
where $\omega_k=\sqrt{{\ibf k}^2+m^2}$ corresponds to the former mass.
We then see that if we use the $\omega-$type free propagators, the result
corresponds to the new dynamics (through the change of pole) and
to the former statistics (corresponding to the mass $m$),
as announced. If, on the contrary, we use the $k_0-$type
free propagators, the resultant statistical distribution will be
$n_B(\sqrt{{\ibf k}^2+m^2+\mu^2})$, \ie the result takes into account
the change of mass also in the statistics.

\section{Conclusion}

On the simple example treated in this part, we have seen that the
vertical part does contribute to the final result, and that
taking into account this contribution is equivalent to the
so-called $n(|k_0|)$ prescription.
Moreover, we insist again on the fact that this contribution is independent
of the times $t_I$ and $t_F$.
We will extend later this property
to a general two point function.

\chapter{A general proof of the $\cmbxXIV t_I-$invariance}

\section{Proof in the case of scalar fields}

In the preceding section, we noticed on an example that the result of
the time integration along the complete path $C$ was independent of
$t_I$, and was invariant by time translation. Before
going on to a general proof of this result, we must emphasize the
fact that this is in complete agreement with our hypothesis of
thermal equilibrium. Indeed, a system in equilibrium
must be invariant by time translation and can't have memory of its initial
time: therefore its Green functions must verify also
these two properties.
We must also insist on their non obvious character:
indeed, the free propagators verify them, but the integration path
does not\footnot{As we will see later, we need an extra property
of the propagators expliciting the fact that they are equilibrium
propagators: \ie the KMS relations.}.

In the following, we will denote  the value of a $n$ point arbitrary
graph by:
$$G_{t_I}(t_1,\cdots ,t_{n})\equiv
\int\limits_{C_{t_I}}^{} \prod\limits_{i=1}^{v} dt_{n+i}
\;F(t_1,\cdots ,t_{n};t_{n+1},\cdots ,t_{n+v}) \no $$
where $v$ is the number of vertices in the graph, and $F(\cdot)$ the
product of the free propagators corresponding to the lines of the
graph, before we perform the vertex-time integrations\footnot{We have not
written the integrations over the three momenta of
 independent loops, because the spatial dependence of
the Green functions has nothing to do with the properties we
are looking at here.}. The $t_I$ index
is there to recall that the integrations are performed on a path
beginning at $t_I$ and ending at $t_I-i\beta$.

Firstly, we can prove the equivalence between the following two properties:
$$\forall (t_1,\cdots ,t_n), \qquad G_{t_I}(t_1,\cdots
 ,t_{n})=G_{t_I}(t_1+a,\cdots ,t_{n}+a)
 \no$$
$$\forall (t_1,\cdots ,t_n),
 \qquad G_{t_I}(t_1,\cdots ,t_{n})=G_{t_I-a}(t_1,\cdots
,t_{n}) \no$$
because the function $F(\cdot)$ depends only on time-differences (see $(14)$).
Therefore, the invariance by time translation is equivalent to the independence
with respect to the initial time $t_I$. We now prove this last property,
first for scalar fields and then for fermions, eventually with a
chemical potential.

For that purpose, we need the following two lemmas:
$${{d\hfil}\over{d t_I}} \int\limits_{C_{t_I}}^{} f(t)\, dt=f(t_I-i\beta)-
f(t_I) \no $$
where $f$ is a function defined on the path $C_{t_I}$, which does
not depend explicitly on $t_I$, and:
$$\left\{\eqalign{&G_0(t_I-i\beta ,t_2;\omega_k)=G_0(t_I,t_2;\omega_k) \cr
&G_0(t_1,t_I-i\beta;\omega_k)=G_0(t_1,t_I;\omega_k) \cr
}\right. \no $$
The last one is only a consequence of the KMS relations, that is to say
of the thermal e\-qui\-li\-brium pro\-per\-ty of the system.

Then, the $F(\cdot)$ function verify also this kind of KMS property:
$$\forall j\in [1,\cdots ,v],\quad
F(\cdots ,t_{n+j-1},t_I-i\beta  ,t_{n+j+1}
,\cdots )=F(\cdots ,t_{n+j-1},t_I
,t_{n+j+1}
,\cdots) \no $$
It is therefore easy to verify by recursion that the $v$
integrations will lead to a function $G_{t_I}(\cdot)$ which does not
depend explicitly on the initial time $t_I$.

\section{Extension to fermions and chemical potential}

The case of fermions with chemical potential is a bit more intricate
because the KMS relations are now:
$$\left\{\eqalign{
&S_0(t_1,t_I;\omega_k)=-e^{-\beta\mu q}S_0(t_1,t_I-i\beta;\omega_k)\cr
&S_0(t_I,t_2;\omega_k)=-e^{\beta\mu q}S_0(t_I-i\beta ,t_2;\omega_k)\cr
}\right. \no$$
where $\mu$ is the chemical potential and $q$ the conserved charge
carried through the propagator from its first point to its second point.
Because the number of fermionic lines going to a vertex is even, the minus
signs cancel mutually when we try to establish $(28)$
with fermions. In the same way, the $\exp(\pm\beta\mu q)$ factors cancel
because of the nullity of the total {\sl conserved\/} charge arriving at a
vertex. So, $(28)$ is still true in the general case of fermions
with chemical potential (of course, the KMS relations
with respect to the external times depend on the type of the external fields),
and that was the only property of the $F(\cdot)$ function we need to
prove the invariance of $G_{t_I}(\cdot)$ with respect to $t_I$.

\section{Application to ITF}

This paragraph lies a bit outside the general line of this article: it
shows how the preceding arguments can lead to define "ITF-like"
Feynman rules, not only on the usual imaginary time path, but also on
every other path going from $t_I$ to $t_I-i\beta$.
The property of $t_I-$invariance can be used to introduce the discrete
Matsubara frequencies of the imaginary time formalism. Let's assume that
we have performed the calculation of a graph in time coordinates: we are
led to a function $G(t_1,\cdots , t_n)$ which does not depend on $t_I$
and which verify KMS relations according to the type of fields
attached to the external lines. The ITF corresponding function can
be defined as \cite{baier3}:
$${\eulr G}(\omega_1,\cdots ,\omega_n)\equiv
\int\limits_{C_{t_I}}^{}\prod\limits_{i=1}^{n}\;
\left(dt_i\;e^{-\omega_it_i}\right)\;G(t_1,\cdots ,t_n) \no $$
Naturally, we require that this ITF function should be independant
of the initial time $t_I$. It is this requirement that will lead us to the
discretization of the Matsubara frequencies.
Let's denote:
$$F(t_1,\cdots ,t_n)\equiv \left(\prod\limits_{i=1}^{n} \,
e^{-\omega_it_i}\right)\; G(t_1,\cdots ,t_n) \no$$
According to the proof given in the preceding paragraphs, the ITF function
will be independent of $t_I$ if the function $F(\cdot)$ verify scalar-like
KMS relations:
$$\forall j\in[1,\cdots ,n],\qquad F(\cdots ,t_{j-1},t_I,t_{j+1},\cdots)=
F(\cdots ,t_{j-1},t_I-i\beta,t_{j+1},\cdots) \no$$
This lead us to the usual frequencies:
\item {(i)} if the line n$^{\circ}$i carries a real scalar:
$$\omega_i={{2\pi n_i}\over{\beta}}\qquad {\rm with}\qquad n_i\in {\bf Z}
 \no$$
\item {(ii)} if the line n$^{\circ}$i carries a fermion:
$$\omega_i={{(2n_i+1)\pi}\over \beta}\qquad {\rm with}\qquad n_i\in {\bf Z}
 \no$$
\item {(iii)} if the line n$^{\circ}$i
carries a fermion with conserved charge $q_i$:
$$\omega_i={{(2n_i+1)\pi}\over \beta}-i\mu q_i\qquad {\rm with} \qquad
n_i\in {\bf Z} \no $$

It is then easy to prove the following property \cite{baier3}:
$${\eulr G}(\omega_1,\cdots ,\omega_n)=-i\beta\delta_{\Sigma,0} \widetilde
{\eulr G}(\omega_2,
\cdots ,\omega_n) \no$$
with
$$\widetilde {\eulr G}(\omega_2,\cdots ,\omega_n)=\int\limits_{C_{t_I}}^{}\;
\prod\limits_{i=2}^{n}\left(dt_i\,e^{-\omega_it_i}\right)\;
G(0,t_2,\cdots ,t_n) \no$$
where $\Sigma=\omega_1+\cdots +\omega_n$ is the total energy entering the
graph.
We can note that the invariance of the result of time integrations
with respect to $t_I$ is
essential in this proof. This is quite normal because the $t_I-$invariance is
related to the invariance by time translation, which is itself related to
the conservation of total energy we want to prove here.

The inverse transformation is:
$$G(t_1,\cdots ,t_n)=\left({i\over \beta}\right)^n\;
\sum\limits_{n_1\cdots n_n\in Z}^{}\;\left(
\prod\limits_{i=1}^{n}\,e^{\omega_it_i}\right)\;
{\eulr G}(\omega_1,\cdots ,\omega_n) \no$$
and we have the following properties:
$$H(t_1,t_2)=\int\limits_{C_{t_I}}^{}dt\;F(t_1,t)G(t,t_2)\Leftrightarrow
{\eulr H}(\omega_1,\omega_2)={i\over\beta}\sum\limits_{n\in
 {\bf Z}}^{}{\eulr F}(\omega_1,\omega)
{\eulr G}(-\omega,\omega_2) \no$$
$$H(t_1,t_2)=F(t_1,t_2)G(t_1,t_2)\Leftrightarrow
{\eulr H}(\omega_1,\omega_2)=
\left({i\over\beta}\right)^2\sum\limits_{n'_1,n'_2\in {\bf Z}}^{}
{\eulr F}(\omega_1',\omega_2'){\eulr G}
(\omega_1-\omega_1',\omega_2-\omega_2') \no$$
The transformed free scalar propagator is:
$${\eulr D}_0(\omega_1,\omega_2)={\beta\over i}\delta_{\omega_1+\omega_2,0}
{1\over{(i\omega_1)^2-\omega_k^2}} \no$$

All these relations enable us to use the usual "ITF" Feynman rules. The
important point to note here is the fact that these Feynman rules are not
at all dependant of the choice of the path going from $t_I$ to $t_I-i\beta$,
so that the name "Imaginary Time Formalism" gives a very restrictive view of
 this
formalism.
\section{Conclusion}

The conclusion of this part is that every graph evaluated on the complete
path $C$ is invariant by time translation and does not depend explicitly
on the initial time $t_I$. In the proof, this property has been shown
to be closely related with thermal equilibrium through the KMS relations.
As a corollary of this result, we can look with suspicion the arguments
based on the limits $t_I\rightarrow -\infty$, $t_F\rightarrow +\infty$
because nothing special occurs when we look at these limits.

\chapter{The vertical part contribution for two point functions}

\section{Spectral representation of two points functions}

In order to extend the result of part \uppercase\expandafter{\romannumeral 3}
to general two point functions, we use their spectral representation.
Let ${\cal F}(t_1,t_2;{\ibf k})$ be a two point function; we can write it as%
 \voir({lands},{nieg}):
$${\cal F}(t_1,t_2;{\ibf k})=\int\limits_{0}^{+\infty}dE\;f(E,
{\ibf k})G_0(t_1,t_2;
E) \no $$
where $f(E,{\ibf k})$ is its spectral density and $G_0(\cdot)$ the free
propagator.
In particular, this is true for a self-energy function:
$$\Sigma(t_1,t_2;{\ibf k})=\int\limits_{0}^{+\infty}dE\;\sigma(E,
{\ibf k})G_0(t_1,t_2;
E) \no $$
Then, if $\Sigma_{ij}$ is the matrix self-energy, we have:
$$\left\{\eqalign{
&\Sigma_{11}(k_0,{\ibf k})=\int\limits_{0}^{+\infty}\!dE\;\sigma(E,{\ibf k}
) D_{11}(k_0,
E)\quad
\Sigma_{22}(k_0,{\ibf k})=\int\limits_{0}^{+\infty}\!dE\;\sigma(E,{\ibf k}
) D_{22}(k_0,
E)\cr
&\Sigma_{12}(k_0,{\ibf k})=-\!\int\limits_{0}^{+\infty}\!dE\;\sigma(E,{\ibf k}
) D_{12}(
k_0,E)\quad
\Sigma_{21}(k_0,{\ibf k})=-\!\int\limits_{0}^{+\infty}\!dE\;\sigma(E,{\ibf k}
) D_{21}(
k_0,E)\cr}\right. \no $$
where the minus signs in the $12$ and $21$ components are due to the
opposite sign of the type $2$ vertices.

\section{The insertion of a self-energy term}

\subsection{Calculation in time coordinates}

We now consider the insertion of a self-energy type function
between two free propagators\footnot{This operation is
the basis of the construction of the complete propagator.} (figure 4):
$$F(t_1,t_2,{\ibf k})\equiv\int\limits_{C}^{}dt_3\,dt_4\;
G_0(t_1,t_3;\omega_k)(-i\Sigma(t_3,t_4;{\ibf k}))G_0(t_4,t_2;\omega_k) \no$$
Thanks to the spectral representation of $\Sigma(\cdot)$, this calculation
is simply the convolution of three free propagators. In the following,
we need only the contribution of the vertical part, which is:
$$F_v(t_1,t_2;{\ibf k})={\beta\over i}{{n_B(\omega_k)(1+n_B(\omega_k))}\over
{(2\omega_k)^2}}\sum\limits_{\epsilon,\eta=\pm 1}^{}e^{-i\eta\omega_k(t_2-
t_1)}\int\limits_{0}^{+\infty}dE\;
{{\epsilon\sigma(E,{\ibf k})}\over{2E}}{{\cal P}\over
{\eta\omega_k-\epsilon E}} \no $$
Its Fourier transformation is independant of the $ij$ component considered
because $F_v(\cdot)$ does not contain any $\theta_c(\cdot)$ function. We have:
$$F_{v,ij}(k_0,{\ibf k})={\beta \over i}
{{n_B(\omega_k)(1+n_B(\omega_k))}\over {2\omega_k}}2\pi\delta(k_0^2-
\omega_k^2)\int\limits_{0}^{+\infty}dE\;\sigma(E,{\ibf k}
){{\cal P}\over{k_0^2-E^2}}
\no $$

\subsection{Calculation in RTF}

Now, we do the same calculation with the usual Feynman rules
of RTF in order to see the difference
between the $n_B(\omega_k)$ and the $n_B(|k_0|)$ prescriptions. We have
to evaluate:
$$F_{\omega[k_0],ij}(k_0,{\ibf k})
\equiv\sum\limits_{a,b=1,2}^{} D_{\omega[k_0],ia}(k_0,
\omega_k)
(-i\Sigma_{ab}(k_0,{\ibf k}))
D_{\omega[k_0],bj}(k_0,\omega_k) \no $$
If we explicit a bit more this quantity, we see that the ill defined
products of distributions cancels mutually thanks to the following
identities:
$$\left\{\eqalign{
&\Sigma_{11}+\Sigma_{22}+\Sigma_{12}+\Sigma_{21}=0\cr
&(\theta(-k_0)+n_B(|k_0|))\Sigma_{21}(k_0,{\ibf k})=
(\theta(k_0)+n_B(|k_0|))\Sigma_{12}(k_0,{\ibf k}) \cr}\right. \no $$
(The first one is always true, and the second one, being a form of the
KMS relation, is only true for equilibrium).

Moreover, this cancellation makes derivatives of $\delta(\cdot)$ functions
appear. For exemple, the $11$ component is:
$$\eqalign{
F_{\omega[k_0],11}&(k_0,{\ibf k})={{i{\cal P}}\over {(k^2-m^2)^2}}\Sigma_{11}
(k_0,{\ibf k})
\cr
&-\pi\left[(1+2n_B)\Sigma_{11}+
(\theta(-k_0)+n_B)\Sigma_{21}+
(\theta(k_0)+n_B)\Sigma_{12}\right]{{\partial\hfil}\over
{\partial k_0^2}}\delta(k_0^2-\omega_k^2)\cr
}\no$$

Therefore, the difference $F_{k_0,ij}(k_0,{\ibf k})-
F_{\omega,ij}(k_0,{\ibf k})$ will not be zero (see appendix 3). By an explicit
calculation, we verify that this difference is independant of the
component $ij$ one is considering:
$$F_{k_0,ij}(k_0,{\ibf k})-F_{\omega,ij}(k_0,{\ibf k})=
{\beta\over i}{{n_B(\omega_k)(1+n_B(\omega_k))}\over{2\omega_k}}
2\pi \delta(k_0^2-\omega_k^2)\left[-i\left(
\Sigma_{11}+{{\Sigma_{12}+\Sigma_{21}
}\over 2}\right)\right] \no $$

\section{Epilogue}

Then, thanks to the spectral representation of the self-energy
matrix and by using the explicit form of the free matrix propagator (see
appendix 2),
we can verify the identity:
$$\Sigma_{11}+{{\Sigma_{12}+\Sigma_{21}}\over 2}=
{{\Sigma_{11}-\Sigma_{22}}\over 2}=\int\limits_{0}^{+\infty}
dE\;\sigma(E,{\ibf k}) {{i{\cal P}}\over {k_0^2-E^2}} \no $$
So that we have:
$$F_{v,ij}(k_0,{\ibf k})=F_{k_0,ij}(k_0,{\ibf k})-F_{\omega,ij}(k_0,{\ibf k})
\no $$
So, we have proven that the contribution of the vertical part corresponds
exactly to the difference between the results given by the $n_B(|k_0|)$
and the $n_B(\omega_k)$ prescriptions.

Therefore, to take this vertical part into account, we have to apply
the $n_B(|k_0|)$ prescription in the usual Real Time Feynman rules\footnot{
Here, we admit that the contribution of the two horizontal parts
is obtained by the usual RTF Feynman rules with $n_B(\omega_k)$ prescription.
This point is rather obvious because the propagators in time coordinates
already contains $n_B(\omega_k)$ distribution functions.}, at
least for two point functions.

\chapter{General remarks on ${\cmbxXIV n}$ point graphs}

\section{Preliminaries}

Before going to the case of higher $n$ point functions, let's consider
the following integral\footnot{This is the kind of calculation we have
to perform in time coordinates, if we except
the three momentum integrations in loops.}:
$${\cal F}(\Sigma)=\int\limits_{C}^{}dt\,F(t) e^{-i\Sigma t} \no$$
where $\Sigma\in {\bf R}$ and where $F(\cdot)$ is the characteristic function
of a subset ${\cal A}$ of $C$ ($F(t)=0$ if $t\notin {\cal A}$, $F(t)=1$ if
$t\in {\cal A}$), with ${\cal A}\cap C_v=\emptyset$ or $C_v$. We will explain
later the reason of this condition. Our purpose is here to determine the
contribution of the vertical part in ${\cal F}(\cdot)$. If ${\cal A}\cap C_v=
\emptyset$, this contribution is trivially zero, so that we only
consider the other case. We have now to distinguish between $\Sigma\not = 0$
and $\Sigma =0$:

(i)  if $\Sigma\not = 0$, we have:
$${\cal F}(\Sigma)={i\over \Sigma}\,e^{-i\Sigma t_I}\,(e^{-\Sigma\beta}-1)+\!
\int\limits_{C_1\oplus C_2}^{}\!dt\, F(t) e^{-i\Sigma t} \no$$
But we know that the result of the calculation of the graph, which is
a sum of such terms, is independent of $t_I$. So, there must
exist other terms which cancel the contribution of the vertical part
in (55), because it depends on $t_I$. The conclusion is that when $\Sigma\not
=0$, (54) will not give a vertical part contribution to the graph.

(ii) if $\Sigma=0$, we have:
$${\cal F}(\Sigma)=-i\beta+\!
\int\limits_{C_1\oplus C_2}^{}\!dt\, F(t) e^{-i\Sigma t} \no$$
Now, the preceding argument cannot be applied because the contribution
of the vertical part does not depend on $t_I$. So, in general, there
is a contribution of the vertical part in (54) to the graph when $\Sigma=0$.

\section{Application to $\cmbxXIII n$ point functions}

The calculation of the vertical part contribution to
the time dependence of a $n$ point function can be reduced
to a sum of integrals like (54), where $F(\cdot)$ is a product of
$\theta_c(\cdot)$ functions and $\Sigma$ a linear combination of
the energies $\omega_i=\sqrt{m^2+{\ibf k}_i^2}$ carried by the various lines of
the graph. The hypothesis made above on ${\cal A}\cap C_v$ is
related to the fact that we have defined the contribution of the vertical
part  to be the one obtained when the integration time is on $C_v$ and the
other times on $C_1\oplus C_2$ (see note 5).

So, we have shown that the result of a vertex integration yields a vertical
part contribution if and only if its $\Sigma$ is zero. Moreover, this
contribution is always finite ($-i\beta$ multiplied by a regular function of
the other times). Then, if we take into account the fact that we have to
perform integrations over the internal independant three momenta, this
 contribution will vanish at the end when $\Sigma$ contains an internal
momentum because $\Sigma=0$ defines a null mesure subset of the set
in which this momentum can vary.

The only remaining possibility to have contributions of the vertical
part is then in terms for which $\Sigma$ contains only the {\sl external\/}
three momenta, or if the diagram contains a self-energy subdiagram (this
situation has been studied in the preceding section).

\section{Example}

In order to illustrate this, we consider the example of figure $5$.
The
external propagators, carrying energy $\omega_i$, are denoted for example:
$$G_0(t_1,u_1;\omega_1)={1\over{2\omega_1}}\sum\limits_{\epsilon_1=\pm 1}^{}
[\theta_c(\epsilon_1(u_1-t_1))+n_B(\omega_1)]e^{-i\epsilon_1
\omega_1(u_1-t_1)} \no$$
and the internal ones are denoted for example:
$$G_0(u_1,u_2;\Omega_1)={1\over{2\Omega_1}}\sum\limits_{\eta_1=\pm 1}^{}
[\theta_c(\eta_1(u_2-u_1))+n_B(\Omega_1)]e^{-i\eta_1\Omega_1(u_2-u_1)} \no$$
So, we have to perform an integral like:
$$\int\limits_{C}^{}\!\prod\limits_{i=1}^{5}\left(du_i e^{-i\Sigma_i
u_i}\right)
\,F(t_1,t_2,t_3;u_1,u_2,u_3,u_4,u_5)
\no$$
with:
$$\left\{\eqalign{
&\Sigma_1=\epsilon_1\omega_1-\eta_1\Omega_1+\eta_6\Omega_6\, ,\quad
\Sigma_2=\epsilon_2\omega_2-\eta_2\Omega_2+\eta_1\Omega_1\, ,\quad
\Sigma_3=\epsilon_3\omega_3-\eta_3\Omega_3+\eta_2\Omega_2\cr
&\Sigma_4=\eta_3\Omega_3-\eta_4\Omega_4-\eta_5\Omega_5\, ,\quad
\Sigma_5=\eta_4\Omega_4+\eta_5\Omega_5-\eta_6\Omega_6\cr
}\right. \no$$
At the first integration (over $u_1$), the $\Sigma$ introduced above is
$\Sigma_1$. At the second integration, it can be $\Sigma_2$ or $\Sigma_1+
\Sigma_2$ because the first integration can generate a term in $\exp(
-i\Sigma_1 u_2)$ . At the third integration, it can be $\Sigma_3$, $\Sigma_1+
\Sigma_3$, $\Sigma_2+\Sigma_3$ or $\Sigma_1+\Sigma_2+\Sigma_3$, and so on.
Among all the possible $\Sigma$ encountered in the five integrations, only
one depends only on the external energies $\omega_i$:
$$\Sigma_1+\Sigma_2+\Sigma_3+\Sigma_4+\Sigma_5=\sum\limits_{i=1}^{3}
\epsilon_i\omega_i \no$$
which is encountered in the last integration. Moreover, because of
the conservation of three momentum, we have $\Omega_3=\Omega_6$ so that:
$$\Sigma_4+\Sigma_5=(\eta_3-\eta_6)\Omega_3 \no$$
and can be zero independantly of $\Omega_3$ when $\eta_3=\eta_6$: this is
precisely what occurs when a subdiagram is of self-energy type.

\section{Conclusion}

The conclusion of this section is that it is possible to have a contribution
of the vertical part of the path in $n>2$ point functions for certain
configurations of the external momenta (namely, when a combination
$\sum\nolimits_{i=1}^{i=n}\epsilon_i\omega_i$ of the external energies
is zero) or when the graph contains a self-energy subdiagram.

\chapter{Concluding remarks}

The main result of this paper is to explicit the role of the vertical
part of the real time path, giving a rather physical interpretation of its
role.
We proved that the vertical part is necessary
in order to have the cancellation of the terms which are not invariant by
time translation, and that the remaining terms do not depend on the
initial time choosen for the path. We must emphasize that such a property
was to be expected on physical considerations  for an equilibrium system.
As a consequence, the usual proofs of the Fourier
Feynman rules based on the limits $t_I\rightarrow
-\infty$, $t_F\rightarrow +\infty$ are thought to be false.

The other result is that the vertical part of the path generally
contributes to the Green functions, as already seen in \cite{nieg}. Moreover,
we have shown in the case of two points functions that this contribution
is simply taken into account by the $n_B(|k_0|)$ prescription associated
to the usual Real Time Feynman rules. This proof is consistent with
the fact that the Green functions does not depend on the
contour parameters, because it works even if we keep $t_I$ finite.

It remains to study more precisely
 the case of the functions with $n>2$ points, and to see if the contribution
of the vertical part which can occur can be dealt simply by the
$n_B(|k_0|)$ prescription. Another problem is
to make more precise the relation between
pathologies and the fact that the
theory is or is not an equilibrium one, as outlined in \voir({alt94a},%
{alt94b}).
\vskip 1cm
\centerline{\bf Aknowledgements}

I would like to thank P. Aurenche and F. Guerin for helpful discussions.
Moreover, I am particularly indebted
to T. Altherr who introduced me
to the domain of finite temperature field theories.
\vskip 1cm
\centerline{{\bf Note added in proof:\hfill}}
After this work had been completed, we became aware of a preprint by
T.S. Evans and A.C. Pearson \cite{ev5}, which contains a rather
different approach to the problem of the vertical parts
of the path in RTF.

In \cite{ev5}, an $\epsilon-$regularized
propagator is used:
$$G_0^{\epsilon}(t_1,t_2;\omega_k)\equiv G_0(t_1,t_2;\omega_k)e^{-\epsilon|
t_2-t_1|} \no $$
which vanishes when the time difference goes to infinity, in association
with the limit $t_I\rightarrow -\infty$. In  order to clarify the
contradiction between their result and ours concerning the
contribution of the vertical part, we repeated the simple
calculation presented in section {\uppercase\expandafter{\romannumeral 3}}
of the present paper with their regularized propagator,
and we were led to the conclusion that the limits $\epsilon\rightarrow 0^{+}$
and $t_I\rightarrow -\infty$ do not commute because of terms like
$\exp(\epsilon\, t_I)$. More precisely:

(i) If we do $\epsilon\rightarrow 0^{+}$ first (this is
equivalent to the use of non-regularized propagators), the result is
invariant by time translation and does
not depend anymore on $t_I$, so that the limit $t_I\rightarrow -\infty$ is
useless. Moreover, there remains a contribution of the vertical
part of the path in the result.

(ii) If we do $t_I\rightarrow -\infty$ first, like in \cite{ev5},
the result is also invariant
by time translation (but this time for different reasons, because the
free regularized propagator does not verify KMS) but the
contribution of the vertical part is lost.

Because of the physical interpretation of thermal equilibrium, we believe
that in a consistent answer to this problem the limit
$t_I\rightarrow -\infty$ has no role to play, so that in time coordinates,
 the use
of non-regularized propagators seems preferable.

\zero

\annexe{The generating functional}

Our purpose is to transform expressions like:
$$\openup -2mm\eqalign{Z^{\omega}[j]=&\omega\left({\rm T}_c
\exp i \int\limits_{\gamma}^{} d^4x
\; {\cal L}_I(\phii(x))+j(x)\phii(x)\right)\cr
=&\exp i\int\limits_{\gamma}^{}d^4x\;
{\cal L}_I\left({{\partial\hfil}\over
{i\partial j(x)}}\right)\;\omega\left({\rm T}_c\exp i\int\limits_{\gamma}^{}
d^4x \; j(x)\phii(x)\right)\cr
} \no $$
To achieve that, we will use twice the Campbell-Haussdorf formula
in order to transform the path ordering in normal ordering \cite{itz}. By
doing this, we make a c-number appear, and the main point is that
a c-number is equal to its average $c=\omega_{1}(c)$, where
$\omega_{1}(\cdot)$ is another arbitrary average operator. The general
result is then (in the case of scalar fields):
$$Z^{\omega}[j]=\exp i\int\limits_{\gamma}^{}d^4x\;
{\cal L}_I\left({{\partial\hfil}\over{i\partial j(x)}}\right)\;
\times\;
Z^{\omega}_0[j] \no$$
with\
$$Z^{\omega}_0[j]=C[j]\times \exp -{1\over
2}\int\limits_{\gamma}^{}d^4x\,d^4y\;
j(x)j(y)\,\omega_1({\rm T}_c \phii(x)\phii(y)) \no$$
$$\openup -2mm\eqalign{C[j]=
\omega\left(
\vphantom{\int}\right.
:\exp i\int\limits_{\gamma}^{}& d^4x\;j(x)\phii(x) :\left.\vphantom{\int}
\right)\cr
&\times\exp{1\over 2}\int\limits_{\gamma}^{}d^4x\,d^4y\;j(x)j(y)
\omega_{1}\left(
\vphantom{\phii^{+}}\right.
\phii^{+}(x)\phii^{-}(y)+\phii^{+}(y)\phii^{-}(x)\cr
&\qquad\qquad\qquad\qquad
{}+\phii^{+}(x)\phii^{+}(y)+\phii^{-}(x)\phii^{-}(y)
\left.\vphantom{\phii^{+}}\right)\cr} \no $$
where $\phii^{+}$ and $\phii^{-}$ are respectively the creation and
the annihilation part of the free field\footnot{This expression for
$Z^{\omega}_0[j]$ is consistent with the Dyson-Schwinger equation
for real scalar fields \cite{lands}:
$$(\dalemb_x+m^2){{\delta Z^{\omega}_0[j]}\over{i\delta j(x)}}=j(x)Z^{
\omega}_0[j] $$}. The first factor in $A2$ will be
expanded in powers of the coupling constant in order to perform the
perturbative expansion. The factor $C[j]$ contains the
information relative to the initial correlations of the system:
we have $C[j]=1$ when the Wick's theorem holds.
The equilibrium field theory is a bit particular because we can
replace $\omh(\cdot)$ by $\omho(\cdot)$ in $C[j]$
 provided we add the vertical part $C_v$ to the path $C_1\oplus C_2$. Then,
by choosing $\omega_{1}=\omho$\footnot{The fact that we can
make $C[j]=1$ with $\omega_1=\omho$ is very important
to have a consistent free theory. Indeed, $\omho({\rm T}_c
\phii(x)\phii(y))$ is a free quantity both
for the dynamics of the fields and for the density operator with which
the average is performed.}, we can make the correlation terms
disappear\footnot{The denomination "correlation terms" is a bit ambiguous
in this context because they depend on the the actual choice
of the integration path  and of the auxiliary average $\omega_1(\cdot)$.
}, so that the Feynman rules will be
the usual ones, the only difference being the integration path.

We can add that in the general case, there is no simple choice of
the path and of the auxiliary average allowing $C[j]=1$. The consequence
of this impossibility is that one will have very intricated Feynman rules
with not only propagators, but also correlation functions \voir({hall1},%
{hall},{kukha}).

\annexe{The free RTF propagators}

Here we recall the free matricial propagators we use in RTF with $\sigma=0$:
$$\left\{\eqalign{
&D_{\omega[k_0],11}(k_0,\omega_k)={{i{\cal P}}\over{k_0^2-\omega_k^2}}+\pi
(1+2n_B(\omega_k[|k_0|]))\delta(k_0^2-\omega_k^2)\cr
&D_{\omega[k_0],22}(k_0,\omega_k)={{-i{\cal P}}\over{k_0^2-\omega_k^2}}+\pi
(1+2n_B(\omega_k[|k_0|]))\delta(k_0^2-\omega_k^2)\cr
&D_{\omega[k_0],12}(k_0,\omega_k)=2\pi(\theta(-k_0)+n_B(\omega_k[|k_0|]))
\delta(k_0^2-\omega_k^2)\cr
&D_{\omega[k_0],21}(k_0,\omega_k)=2\pi(\theta(k_0)+n_B(\omega_k[|k_0|]))
\delta(k_0^2-\omega_k^2)\cr}\right. \no$$
where the argument of the distribution functions is $\omega_k$ or $|k_0|$,
according to the prescription we choose.

\annexe{Distributions and pathologies}

The distributions we have to deal with in this paper are
$\delta(x)$ and ${\cal P}/x$. The squares $\delta^2(x)$ and
$({\cal P}/x)^2$ are ill defined, even in terms of distributions,
but it is possible to give a sense to a linear
combination of them. To
do so, the proper method is to regularize the distributions
we are interested in. A simple regularization is given by \cite{lands}:
$$\lim_{\epsilon\rightarrow 0^{+}}
\quad\left[{1\over{x+i\epsilon}}\right]^{n+1}={{\cal P}\over{x^{n+1}}}+
i{{\pi(-1)^{n+1}}\over{n!}}{{d^n\hfil}\over{dx^n}}\delta(x) \no $$
where the derivative is defined in the sense of distributions:
$$\forall f,\qquad D'(f)=-D(f') \no $$
By using $(A5)$ for $n=0$ and $n=1$, we obtain the following
relations:
$${{\cal P}\over{x^2}}=\left({{\cal P}\over x}\right)^2-\pi^2\delta^2(x) \no $$
$$2\delta(x){{\cal P}\over x}=-\delta'(x) \no $$
These relations, combined to the definition of the distributional derivative,
imply that the prescription for the argument in the statistical factors
is important each time we have to multiply free propagators carrying the
same momentum.
Indeed, if we look at a product like $a(x)\delta(x)({\cal P}/x)$, the
result will be a different distribution
if we replace $x$ by $0$ in $a(\cdot)$\footnot{Of course, we have
$a(x)\delta(x)=a(0)\delta(x)$. In fact, the problem comes from the fact
that we cannot define unambiguously an associative multiplication of
distributions.}.
In order to see that\footnot{Another way to see the problem is
to perform the inverse Fourier transformation of a function where
pathologies does arise. Then, one will encounter double poles,
the residue of which contains derivatives: the various
 prescriptions for the arguments of the distribution functions will
 lead us to different results.},
 consider the distribution $(a(x)-a(0))\delta'(x)$ and
apply it to a test function $f(x)$; the result is  not zero:
$$\eqalign{\int\limits_{-\infty}^{+\infty}dx\;(a(x)-a(0))\delta'(x)f(x)&=
-\int\limits_{-\infty}^{+\infty}dx\;\delta(x)[(a(x)-a(0))f(x)]'\cr
&=-a'(0)f(0)\not= 0\cr} \no $$

\vfill\eject

\etext

\hbox {}\vskip 1cm
\centerline {\hbox {{\fam \itbfXIVfont \cmbxXIVf REFERENCES}}}
\vskip 0.8cm
\vskip 2mm
$[1]$\hskip 3mm {\fam \slfam \tensl R.L. Kobes, K.L. Kowalski } \hskip 3mm
Phys. Rev. D {\fam \bffam \tenbf 34}, 513 ($\fam \@ne 1986$).
\vskip 2mm
$[2]$\hskip 3mm {\fam \slfam \tensl R.L. Kobes, G.W. Semenov, N. Weiss } \hskip
3mm Z. Phys. C {\fam \bffam \tenbf 29}, 371 ($\fam \@ne 1985$).
\vskip 2mm
$[3]$\hskip 3mm {\fam \slfam \tensl N.P. Landsman, Ch.G. van Weert } \hskip 3mm
Phys. Rep. {\fam \bffam \tenbf 145}, 141 ($\fam \@ne 1987$).
\vskip 2mm
$[4]$\hskip 3mm {\fam \slfam \tensl M. Marinaro } \hskip 3mm Phys. Rep. {\fam
\bffam \tenbf 137}, 81 ($\fam \@ne 1986$).
\vskip 2mm
$[5]$\hskip 3mm {\fam \slfam \tensl A.J. Niemi, G.W. Semenov } \hskip 3mm Nucl.
Phys. B {\fam \bffam \tenbf 230}, 181 ($\fam \@ne 1984$).
\vskip 2mm
$[6]$\hskip 3mm {\fam \slfam \tensl A.J. Niemi } \hskip 3mm Phys. Lett. B {\fam
\bffam \tenbf 203}, 425 ($\fam \@ne 1987$).
\vskip 2mm
$[7]$\hskip 3mm {\fam \slfam \tensl L. Van Hove } \hskip 3mm Phys. Rep. {\fam
\bffam \tenbf 137}, 11 ($\fam \@ne 1986$).
\vskip 2mm
$[8]$\hskip 3mm {\fam \slfam \tensl T.S. Evans } \hskip 3mm Z. Phys. C {\fam
\bffam \tenbf 36}, 153 ($\fam \@ne 1987$).
\vskip 2mm
$[9]$\hskip 3mm {\fam \slfam \tensl T.S. Evans } \hskip 3mm Z. Phys. C {\fam
\bffam \tenbf 41},333 ($\fam \@ne 1988$).
\vskip 2mm
$[10]$\hskip 3mm {\fam \slfam \tensl A. Niegawa } \hskip 3mm Phys. Rev. D {\fam
\bffam \tenbf 40}, 1199 ($\fam \@ne 1989$).
\vskip 2mm
$[11]$\hskip 3mm {\fam \slfam \tensl T.S. Evans } \hskip 3mm Phys. Rev. D {\fam
\bffam \tenbf 47}, 4196 ($\fam \@ne 1993$).
\vskip 2mm
$[12]$\hskip 3mm {\fam \slfam \tensl P. Danielewicz } \hskip 3mm Ann. of Phys.
{\fam \bffam \tenbf 152}, 239 ($\fam \@ne 1984$).
\vskip 2mm
$[13]$\hskip 3mm {\fam \slfam \tensl J. Rammer, H. Smith } \hskip 3mm Rev. of
Modern Physics {\fam \bffam \tenbf 58}, 323 ($\fam \@ne 1986$).
\vskip 2mm
$[14]$\hskip 3mm {\fam \slfam \tensl C. Itzykson, J.B. Zuber } \hskip 3mm Ed.
Mac Graw Hill
\vskip 2mm
$[15]$\hskip 3mm {\fam \slfam \tensl Y. Fujimoto, H. Matsumoto, H. Umezawa, I.
Ojima } \hskip 3mm Phys. Rev. D {\fam \bffam \tenbf 30}, 1400 ($\fam \@ne
1984$).
\vskip 2mm
$[16]$\hskip 3mm {\fam \slfam \tensl R. Baier, A. Niegawa } \hskip 3mm Phys.
Rev. D {\fam \bffam \tenbf 49}, 4107 ($\fam \@ne 1994$).
\vskip 2mm
$[17]$\hskip 3mm {\fam \slfam \tensl T. Altherr } \hskip 3mm Phys. Lett. B
{\fam \bffam \tenbf 333}, 149 ($\fam \@ne 1994$).
\vskip 2mm
$[18]$\hskip 3mm {\fam \slfam \tensl T. Altherr } \hskip 3mm Preprint
cern-th.7336/94 (to be published in Phys. Lett. B).
\vskip 2mm
$[19]$\hskip 3mm {\fam \slfam \tensl T.S. Evans, A.C. Pearson } \hskip 3mm
Preprint Imperial/TP/93-94/09.
\vskip 2mm
$[20]$\hskip 3mm {\fam \slfam \tensl A.G. Hall } \hskip 3mm Physica A {\fam
\bffam \tenbf 80}, 369 ($\fam \@ne 1975$.
\vskip 2mm
$[21]$\hskip 3mm {\fam \slfam \tensl A.G. Hall } \hskip 3mm J. Phys. A {\fam
\bffam \tenbf 8}, 214 ($\fam \@ne 1975$).
\vskip 2mm
$[22]$\hskip 3mm {\fam \slfam \tensl Y.A. Kukharenko, S.G. Tikhodeev } \hskip
3mm Sov. Phys. JETP {\fam \bffam \tenbf 56}, 831 ($\fam \@ne 1982$).

\def\typepartie{chapitre}
\cfigure=0
\vfill\eject
\null\vfill


\centerline{\cmbxXIV FIGURE CAPTIONS}
\vskip 1cm
\caption{The Real Time Path}
\caption{The Real Time Path with $\sigma=0$}
\caption{Repeated insertions of mass terms}
\caption{Insertion of a self-energy function}
\caption{An example of $3$ point diagram}
\vfill\eject
\cfigure=0
\centerline{\cmbxXIV FIGURES}
\vfill
\xfigure{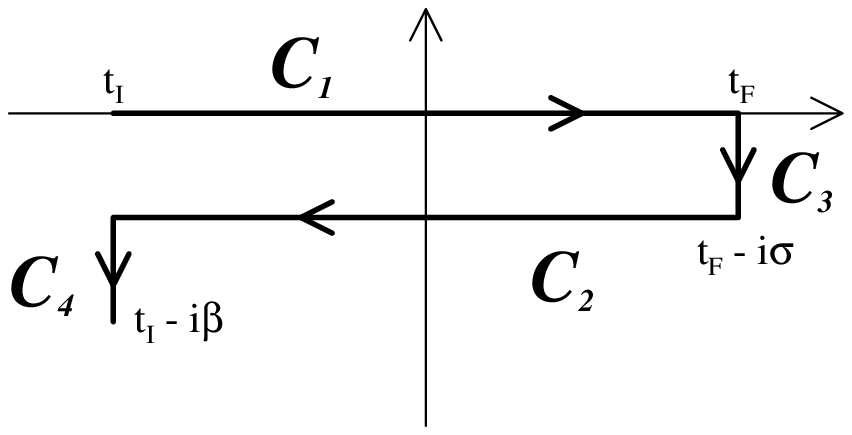}
\vfill
\xfigure{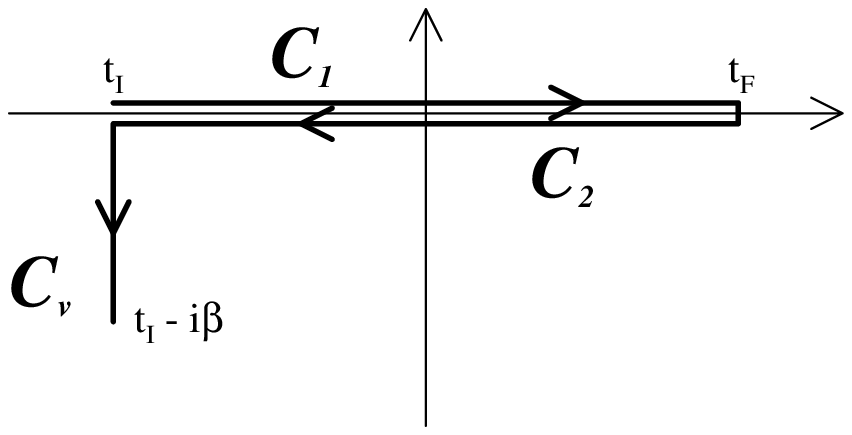}
\vfill\eject
\null\vfill
\xfigure{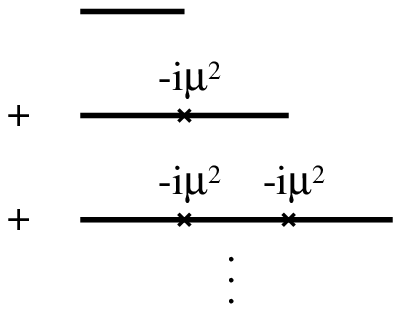}
\vfill
\xfigure{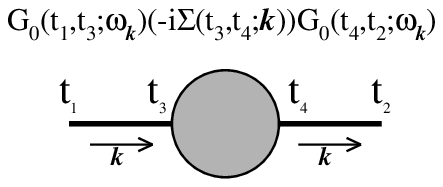}
\vfill
\xfigure{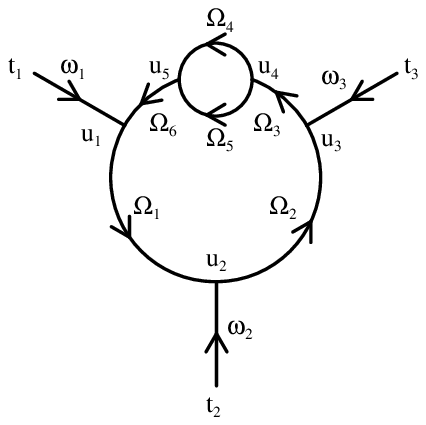}
\vfill\eject

\end